\documentclass[english,aps,pra,reprint,superscriptaddress]{revtex4-1}
\usepackage[T1]{fontenc}
\usepackage[latin9]{inputenc}
\usepackage{color}
\usepackage{babel}
\usepackage{amsmath}
\usepackage{amssymb}
\usepackage{graphicx}
\usepackage{wasysym}
\usepackage[unicode=true,
 bookmarks=true,bookmarksnumbered=false,bookmarksopen=false,
 breaklinks=false,pdfborder={0 0 1},backref=false,colorlinks=true]
 {hyperref}
\hypersetup{
 pdfborderstyle=,allcolors=blue}

\makeatletter

\newcommand*\LyXThinSpace{\,\hspace{0pt}}

\usepackage{babel}
\usepackage{braket}
\usepackage{dsfont}

\makeatother

\begin{document}
\title{Coherence effects in the performance of the quantum Otto heat engine}
\author{Patrice A. Camati}
\email{patrice.camati@neel.cnrs.fr}

\affiliation{Centro de Ciências Naturais e Humanas, Universidade Federal do ABC,
Avenida dos Estados 5001, 09210-580 Santo André, São Paulo, Brazil}
\affiliation{Univ. Grenoble Alpes, CNRS, Grenoble INP, Institut Néel, 38000 Grenoble,
France}
\author{Jonas F. G. Santos}
\email{jonas.floriano@ufabc.edu.br}

\affiliation{Centro de Ciências Naturais e Humanas, Universidade Federal do ABC,
Avenida dos Estados 5001, 09210-580 Santo André, São Paulo, Brazil}
\author{Roberto M. Serra}
\email{serra@ufabc.edu.br}

\affiliation{Centro de Ciências Naturais e Humanas, Universidade Federal do ABC,
Avenida dos Estados 5001, 09210-580 Santo André, São Paulo, Brazil}
\begin{abstract}
The working substance fueling a quantum heat engine may contain coherence
in its energy basis, depending on the dynamics of the engine cycle.
In some models of quantum Otto heat engines, energy coherence has
been associated with entropy production and quantum friction. We considered
a quantum Otto heat engine operating at finite time. Coherence is
generated and the working substance does not reach thermal equilibrium
after interacting with the hot heat reservoir, leaving the working
substance in a state with residual energy coherence. We observe an
interference-like effect between the residual coherence (after the
incomplete thermalization) and the coherence generated in the subsequent
finite-time stroke. We introduce analytical expressions highlighting
the role of coherence and examine how this dynamical interference
effect influences the engine performance. Additionally, in this scenario
in which coherence is present along the cycle, we argue that the careful
tuning of the cycle parameters may exploit this interference effect
and make coherence acts like a dynamical quantum lubricant. To illustrate
this, we numerically consider an experimentally feasible example
and compare the engine performance to the performance of a similar
engine where the residual coherence is completely erased, ruling out the
dynamical interference effect.
\end{abstract}
\maketitle

\section{Introduction}

One of the aims of quantum thermodynamics is to describe, at a fundamental
level, the energy and entropy exchange among systems~\cite{Esposito2009,Kosloff2013,Vinjanampathy2016,Millen2016,Alicki2018}.
The focus on the description and control of small quantum systems
greatly spurred the thermodynamics of quantum heat engines and refrigerators~\cite{Alicki2018,Gelbwaser-Klimovsky2015}.
Experimentally, a single-ion heat engine~\cite{Ro=00003D0000DFnagel2016},
a three-ion refrigerator~\cite{Maslennikov2017}, and an Otto cycle
exploring the harmonic oscillations of a nanobeam~\cite{Klaers2017}
have been recently implemented. Even more recently, a quantum Otto
heat engine employing a spin working substance~\cite{Peterson2018}
and an ensemble of nitrogen-vacancy centers in diamonds~\cite{Klatzow2019}
have been reported. On the other hand, coherence is one of the fundamental
properties of nature, setting apart the quantum from the classical
descriptions of reality. Measures to quantify coherence have been
recently proposed~\cite{Aberg2006,Baumgratz2014,Streltsov2017},
applying similar methods used to quantify entanglement. In particular,
some measures have operational meaning, quantifying the distillation~\cite{Winter2016}
and the erasing cost of quantum coherence~\cite{Singh2017}.

The role of coherence was theoretically addressed employing the photo-Carnot
engine~\cite{Scully2003}, which models the working substance as
a four-level system. The photo-Carnot engine is an extension of the
model employed to thermodynamically describe the laser~\cite{Scovil1959,Geusic1967},
which is fueled by a three-level working substance. These models employ
what could be called a ``partial-spectrum thermalization'' (PST),
in which the heat source interacts only with a subset of the energy
states, thus thermalizing part of the spectrum.

The PST approach to quantum machines has been one of the major frameworks
to analyze the role of coherences in quantum machines~\cite{Scully2003,Scully2010,Scully2011,Dorfman2011,Rahav2012,Dorfman2013,Goswami2013,Uzdin2015,T=00003D0000FCrkpen=00003D0000E7e2016,Uzdin2016,Dorfman2018}.
For instance, employing the approach developed in Ref.~\cite{Uzdin2015},
the recent experiment with nitrogen-vacancy centers in diamonds~\cite{Klatzow2019}
showed the presence of a quantum signature in the power of engines
in the so-called small action limit~\cite{Uzdin2015}. The role of
coherence has also been addressed in other approaches~\cite{Brandner2017,Dodonov2018,Marvian2018}.
Here, we focus on the quantum Otto heat engine (QOHE)~\cite{Kosloff2017}.
The effects of coherence in this engine model, which is different
from the PST model, have been less investigated.

A heat engine does not attain its theoretically maximum efficiency
due to entropy production, the thermodynamic quantifier of irreversibility~\cite{Grootbook1984,Lebonbook2008,=00003D0000C7engelbook2015}.
In classical thermodynamics, two processes are responsible for the
irreversibility of engines. The external friction, or simply friction,
is associated with the exchange of energy at the system boundary due
to sliding. The internal friction~\cite{Rezek2010} is associated
with the finite-time engine operation. It is manifested by the disparity
between the internal dynamics and operation timescales. In order to
achieve the best engine efficiency, the engine should operate quasistatically
and be frictionless, in which case the entropy production is zero
throughout the cycle. However, from the practical point of view, this
mode of operation is not interesting since it would output zero or
very low power.

A new kind of (internal) friction in microscopic engines with quantum
working substances, intrinsically non-classical in nature, has been
studied in the past decades~\cite{Rezek2010,Kosloff2002,Feldmann2003,Feldmann2004,Feldmann2006,Rezek2006,Feldmann2012,Zagoskin2012,Thomas2014,Plastina2014,Alecce2015,Correa2015,Campo2018}.
The origin of such a quantum friction is attributed to the noncommutativity
of the driving Hamiltonian at different times~\cite{Kosloff2002,Feldmann2003,Feldmann2004,Feldmann2006,Rezek2006,Rezek2010,Feldmann2012,Plastina2014,Thomas2014,Alecce2015},
which induces transitions among the instantaneous energy eigenstates.
Furthermore, when operating in the quasistatic regime (transitionless
regime), the quantum friction becomes zero~\cite{Kosloff2002,Feldmann2004,Feldmann2006,Rezek2006,Rezek2010,Feldmann2012,Plastina2014,Thomas2014,Alecce2015},
just as the internal friction of classical engines would. This research
avenue spurred the idea of quantum lubrication, which seeks to render
the effects of quantum friction negligible while operating the quantum
engine at finite time. One of the most commonly employed strategies
is to perform shortcuts to quantum adiabaticity~\cite{Campo2018,Torrontegui2013}.
This method adds so-called counteradiabatic driving fields that make
the working substance evolve in a transitionless dynamics~\cite{Campo2014,Funo2017}.
How costly is such an additional control remains an open question~\cite{Zheng2016,Campbell2017,Calzetta2018}.

On the one hand, some investigations have connected the coherence
in quantum engines to quantum friction~\cite{Feldmann2006,Rezek2006,Feldmann2012,Thomas2014,Correa2015}.
On the other hand, other investigations connected coherence to an
increase in the entropy produced in a thermalization process~\cite{Santos2017,Francica2017}.
However, no simple expression explicitly relating the coherence to
the engine efficiency and power output has been obtained so far.

We present analytical expressions that relate the entropy production
and quantum friction to the coherence in the energy basis (energy
coherence) of the working substance along the cycle. Employing the
relation between the efficiency and power in terms of entropy production
and quantum friction, power and efficiency can be directly linked
to the energy coherence.

Moreover, we considered an incomplete thermalization (second) stroke
after a finite-time driven (first) stroke that generated coherence
in the energy basis. Thus, some of this generated coherence is retained
in the state after the incomplete thermalization stroke (residual
coherence) [see Figs.~\ref{fig:Generic-engine-cycle.}\textbf{(a)}
and \ref{fig:Generic-engine-cycle.}\textbf{(b)}]. We employed a numerical
example which is experimentally feasible to show that the residual
coherence interferes with the coherence generated in the third finite-time
stroke. We define an alternative engine in which a full dephasing
operation in the energy basis completely erases this residual coherence
(the dephased engine) and compare the performance of both engines,
with and without this dynamical interference effect.

We argue that the careful tuning of the cycle parameters (driving
and thermalization times) can make the engine run in the ``constructive
regime,'' where this interference effect can be exploited to enhance
the engine performance when compared to the dephased engine. Therefore,
the interference can be seen as a dynamical quantum lubricant. We
stress that this comparison is not between classical and quantum setups
but rather between two quantum settings.

\section{The Engine Cycle\label{sec:The-Engine-Cycle}}

\begin{figure}[t]
\includegraphics{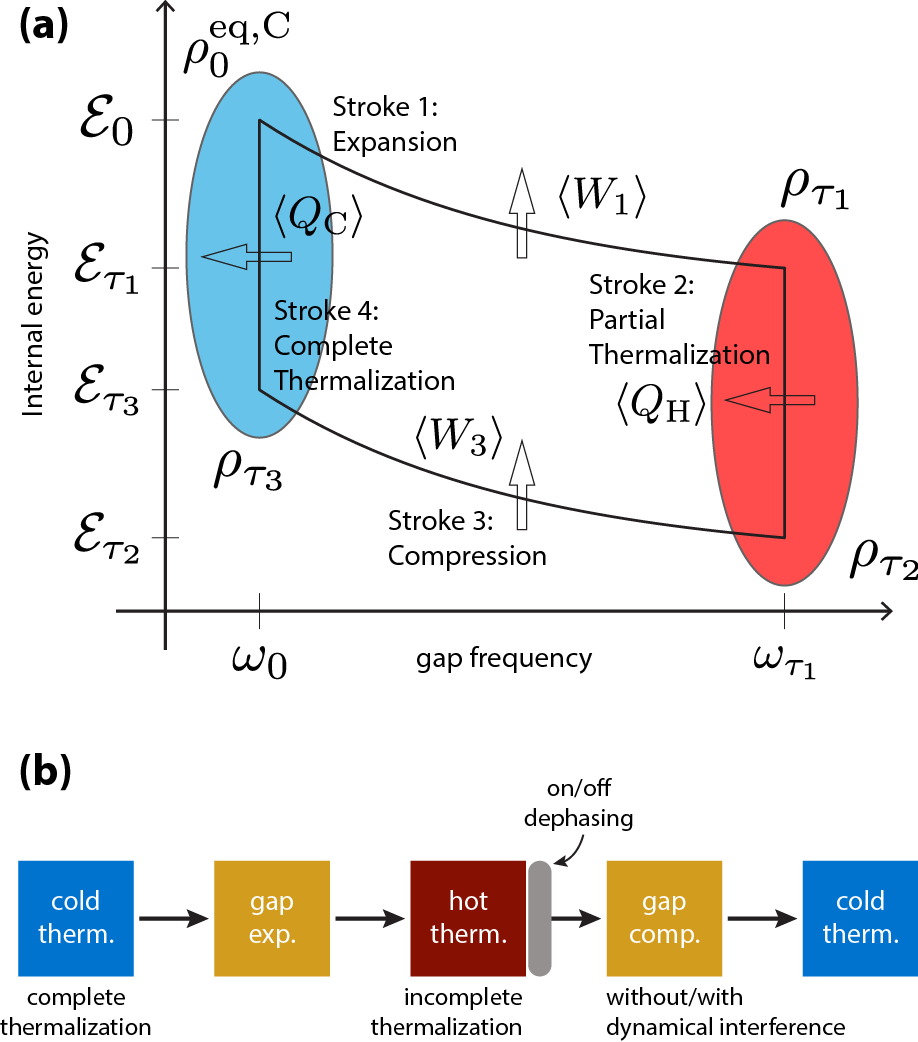}

\caption{Engine cycle. \textbf{(a)} The working substance begins in the cold
thermal state $\rho_{0}^{\text{eq,c}}$, with Hamiltonian $H_{0}$
and inverse temperature $\beta_{\text{c}}$. The working substance
is driven by an adiabatic expansion which changes the Hamiltonian
to $H_{\tau_{1}}$ and leads to the state $\rho_{\tau_{1}}$ at time
$\tau_{1}$. The second stroke is comprised by hot thermalization,
where the working substance interacts with a hot heat reservoir. The
interaction time between the working substance and the heat reservoir
is sufficiently small such that the thermalization is incomplete.
The third stroke is an adiabatic compression changing the Hamiltonian
from $H_{\tau_{1}}$ back to $H_{0}$. The fourth stroke is a complete
thermalization with the cold heat reservoir. \textbf{(b)} Representation
of the cycle with a dephasing operation in the energy basis after
the incomplete thermalization. For this dephased engine, the states
at the end of each stroke are $\rho_{0}^{\text{eq,c}}$, $\rho_{\tau_{1}}$,
$\rho_{\tau_{2}}^{\text{deph}}$, and $\rho_{\tau_{3}}^{\text{deph}}$.
This dephasing operation does not cost energy and erases the residual
coherence of the engine. \label{fig:Generic-engine-cycle.} }
\end{figure}

Let us consider a single-qubit working substance which fuels a QOHE
similar to the system employed in the experimental implementation
of Ref.~\cite{Peterson2018}. The stroke-driven engine cycle is comprised
by two Hamiltonian driven protocols (energy gap expansion and compression)
and two undriven thermalization strokes, which are depicted in Fig.~\ref{fig:Generic-engine-cycle.}\textbf{(a)}.
In Otto engines, the work and heat exchanges are separated among the
strokes: work is only exchanged in the two driven strokes and heat
is only exchanged in the two undriven thermalization strokes.

The working substance begins in the cold Gibbs state $\rho_{0}^{\text{eq,c}}=e^{-\beta_{\text{c}}H_{0}}/Z_{0}^{\text{c}}$,
where $\beta_{\text{c}}=\left(k_{B}T_{\text{c}}\right)^{-1}$ is the
cold inverse temperature, $H^{\text{exp}}\left(0\right)=H_{0}$ is
the initial Hamiltonian (``$\text{exp}$'' stands for expansion),
and $Z_{0}^{\text{c}}=\text{Tr}\left[e^{-\beta_{\text{c}}H_{0}}\right]$
is the associated partition function. The initial Hamiltonian is given
by $H_{0}=\frac{\hbar\omega_{0}}{2}\sigma_{x}$, where $\omega_{0}$
is the initial transition frequency, and $\sigma_{x,y,z}$ denote
the Pauli matrices.

In the first stroke, the energy gap of the working substance is increased
by the driven Hamiltonian $H^{\text{exp}}\left(t\right)$ in a unitary
dynamics. The working substance is assumed to be disconnected from
the heat sources so that no energy is exchanged with them. In a realistic
scenario, one could consider that the driven time is fast enough so
that the energy exchanged between system and environment can be neglected
and the driven dynamics is well described by a unitary evolution~\cite{Peterson2018,Batalh=00003D0000E3o2014,Batalh=00003D0000E3o2015,Camati2016}.
Hence, the state after the expansion stroke is given by $\rho_{\tau_{1}}=U_{\tau_{1},0}\rho_{0}^{\text{eq,c}}U_{\tau_{1},0}^{\dagger}$,
where $U_{\tau_{1},0}=\mathcal{T}_{>}\exp\left\{ -\frac{i}{\hbar}\int_{0}^{\tau_{1}}dt\,H^{\text{exp}}\left(t\right)\right\} $,
$\mathcal{T}_{>}$ is the time-ordering operator, $t\in\left[0,\tau_{1}\right]$,
and 
\begin{equation}
H^{\text{exp}}\left(t\right)=\frac{\hbar\omega\left(t\right)}{2}\left[\cos\left(\frac{\pi t}{2\tau_{1}}\right)\sigma_{x}+\sin\left(\frac{\pi t}{2\tau_{1}}\right)\sigma_{z}\right],\label{eq:driving hamiltonian}
\end{equation}
with $\omega\left(t\right)=\omega_{0}\left(1-\frac{t}{\tau_{1}}\right)+\omega_{\tau_{1}}\left(\frac{t}{\tau_{1}}\right)$.
We have chosen this protocol design because it has been recently employed
in an experimental realization of the quantum Otto cycle with spin
qubits~\cite{Peterson2018}.

In the second stroke, the working substance interacts with a hot heat
reservoir at inverse temperature $\beta_{\text{h}}=\left(k_{B}T_{\text{h}}\right)^{-1}$
and it undergoes a hot thermalization. The Hamiltonian is kept fixed
at $H^{\text{hot}}\left(t\right)=H^{\text{exp}}\left(\tau_{1}\right)=H_{\tau_{1}}=\frac{\hbar\omega_{\tau_{1}}}{2}\sigma_{z}$
spanning the time interval $t\in\left[\tau_{1},\tau_{2}\right]$.
Some stroke-driven models of quantum heat engines assume that this
thermalization stroke is complete so that the system reaches thermal
equilibrium state at the end of the stroke~\cite{Wang2009,Altintas2015}.
More precisely, in order to achieve this complete thermalization the
condition $\tau_{2}-\tau_{1}=\tau_{\text{therm}}^{\text{h}}\gg\tau_{\text{relaxation}}^{\text{h}}$
should be satisfied, where $\tau_{\text{therm}}^{\text{h}}$ is the
thermalization time and $\tau_{\text{relaxation}}^{\text{h}}$ is
the relaxation time of the working substance with the hot heat reservoir.

Since our Hamiltonian does not commute with itself at different times,
the first stroke generates coherence in the energy basis, all of which
would be erased if such a complete thermalization was performed. Therefore,
we consider a incomplete hot thermalization stroke, in which the thermalization
time is of the order $\tau_{\text{therm}}^{\text{h}}\lesssim\tau_{\text{relaxation}}^{\text{h}}$.
Performing an incomplete thermalization in our QOHE model will allow
the working substance to retain a residual amount of coherence at
the end of the hot thermalization (second stroke). Thus, there is
some coherence generated at the first stroke that endures the thermalization
stroke and, hence, will be present at the next driven (third) stroke.
Therefore, the incomplete thermalization allows the dynamical transference
of (some) coherence from the first to the third stroke. One of our
goals is to study how the presence of this residual coherence in the
dynamics of the cycle changes the thermodynamic quantities and the
performance of the engine.

In the third stroke, the working substance energy gap is decreased
to its original value during a unitarily driven dynamics. The compression
Hamiltonian drives the qubit according to the condition $H^{\text{com}}\left(t\right)=H^{\text{exp}}\left(\tau_{1}+\tau_{2}-t\right)$
for the time interval $t\in\left[\tau_{2},\tau_{3}\right]$, with
$\tau_{3}-\tau_{2}=\tau_{1}$ (``$\text{com}$'' stands for compression).
This condition guarantees that $H^{\text{com}}\left(t\right)$ takes
the same values that $H^{\text{exp}}\left(t\right)$ did in the expansion
stroke, but in inverse order (see~\ref{subsec:Relation-between-expansion and compression strokes}
for a detailed explanation). Denoting by $\rho_{\tau_{2}}$ the final
state of the second stroke, the state after the compression stroke
is given by $\rho_{\tau_{3}}=V_{\tau_{3},\tau_{2}}\rho_{\tau_{2}}V_{\tau_{3},\tau_{1}}^{\dagger}$,
where $V_{\tau_{3},\tau_{2}}=\mathcal{T}_{>}\exp\left\{ -\frac{i}{\hbar}\int_{\tau_{2}}^{\tau_{3}}dt\,H^{\text{com}}\left(t\right)\right\} $
.

The fourth stroke is an undriven thermalization with a cold heat reservoir
at inverse temperature $\beta_{\text{c}}$. The stroke spans the time
interval $t\in\left[\tau_{3},\tau_{4}\right]$ and the Hamiltonian
is kept fixed at $H^{\text{cold}}\left(t\right)=H^{\text{com}}\left(\tau_{3}\right)=H^{\text{exp}}\left(0\right)=H_{0}$.
In order to close the engine cycle, i.e., $\rho_{\tau_{4}}=\rho_{0}^{\text{eq,c}}$,
we consider complete thermalization in this stroke. Therefore, the
cold thermalization time must satisfy the condition $\tau_{4}-\tau_{3}=\tau_{\text{therm}}^{\text{c}}\gg\tau_{\text{relaxation}}^{\text{c}}$.

The four relevant energetic quantities to analyze the thermodynamics
of the engine are the following. The first- and third-stroke works
$\left\langle W_{1}\right\rangle =\mathcal{E}_{\tau_{1}}-\mathcal{E}_{0}$
and $\left\langle W_{3}\right\rangle =\mathcal{E}_{\tau_{3}}-\mathcal{E}_{\tau_{2}}$,
respectively, where $\mathcal{E}_{t}=\text{Tr}\left[H\left(t\right)\rho_{t}\right]$
denotes the mean instantaneous internal energy; and the hot and cold
heats $\left\langle Q_{\text{h}}\right\rangle =\mathcal{E}_{\tau_{2}}-\mathcal{E}_{\tau_{1}}$
and $\left\langle Q_{\text{c}}\right\rangle =\mathcal{E}_{0}-\mathcal{E}_{\tau_{3}}$,
which are the energies absorbed by the working substance during the
interaction with the hot and cold heat reservoirs, respectively.

The dynamics of a qubit with a Hamiltonian $H\left(t\right)$ with
energy gap $\hbar\omega$ interacting with a Markovian heat reservoir
at inverse temperature $\beta$ can be described by the master equation
\cite{Breuerbook2003,Chakraborty2016} 
\begin{align}
\frac{d}{dt}\rho\left(t\right)= & -\frac{i}{\hbar}\left[H\left(t\right),\rho\left(t\right)\right]\nonumber \\
 & +\gamma_{\downarrow}\left[\Gamma_{\downarrow}\rho\left(t\right)\Gamma_{\uparrow}^{\dagger}-\frac{1}{2}\left\{ \rho\left(t\right),\Gamma_{\uparrow}^{\dagger}\Gamma_{\downarrow}\right\} \right]\nonumber \\
 & +\gamma_{\uparrow}\left[\Gamma_{\uparrow}\rho\left(t\right)\Gamma_{\downarrow}^{\dagger}-\frac{1}{2}\left\{ \rho\left(t\right),\Gamma_{\downarrow}^{\dagger}\Gamma_{\uparrow}\right\} \right],\label{eq:master equation qubit}
\end{align}
where $\gamma_{\downarrow}=\gamma_{0}\left(N_{\text{BE}}+1\right)$,
$\gamma_{\uparrow}=\gamma_{0}N_{\text{BE}}$, $\gamma_{0}$ is the
vacuum decay rate, $N_{\text{BE}}=\left(e^{\beta\hbar\omega}-1\right)^{-1}$
is the Bose-Einstein distribution, and $\Gamma_{\downarrow}$ ($\Gamma_{\uparrow}$)
is the ladder operator in the energy eigenbasis that takes the excited
(ground) state and transforms it into the ground (excited) state.
The analytical solution of this equation is used to obtain the state
$\rho_{\tau_{2}}$ after the incomplete hot thermalization stroke,
and hence the thermodynamic relations with incomplete thermalization
(for details see \ref{sec:The-Engine-Cycle-1}).

The residual coherence that is transferred from the expansion to the
compression strokes due to incomplete thermalization affects the thermodynamic
quantities. In particular, as we discuss in Sec.~\ref{subsec:Numerical-Analysis},
a dynamical interference effect between the residual coherence and
the coherence generated in the third stroke is revealed. In order
to pinpoint the consequences of this interference effect, we benchmark
our quantum engine with an alternative cycle without dynamical interference.
In such a cycle a dephasing operation (in the energy basis) is employed
to completely erase the residual coherence (after the second stroke),
even for small thermalization times {[}see Fig.~\ref{fig:Generic-engine-cycle.}\textbf{(b)}{]}.
This dephasing operation in the energy basis has no energetic cost,
since it does not change the working substance mean internal energy.

In this way, we have two quantum engines, with and without (the dephased
engine) dynamical interference, providing a fair benchmark for the
investigation of the coherence effects along the cycle. Additionally,
we employ the superscript ``$\text{deph}$'' to describe the quantities
corresponding to the dephased engine cycle {[}see caption of Fig.~\ref{fig:Generic-engine-cycle.}\textbf{(b)}{]}.
Further details concerning the dynamical interference effect are discussed
in Sec.~\ref{subsec:Numerical-Analysis}.

\section{The role of quantum coherence in the irreversibility and performance
of the engine \label{sec:Thermodynamic-Quantities-and}}

\subsection{General description\label{subsec:General-Description}}

The four relevant states $\rho_{0}^{\text{eq,c}}$, $\rho_{\tau_{1}}$,
$\rho_{\tau_{2}}$, and $\rho_{\tau_{3}}$ (related to the four strokes)
are the key states of the engine cycle that will be employed to completely
analyze the performance of the proposed engine. For further reference,
we call this set of states the key-working-substance states.

Before we proceed, it is convenient to establish a few important quantities
that are going to be important throughout our analyzes. The (Kullback-Leibler-Umegaki)
divergence between an arbitrary state $\rho$ and a reference state
$\rho^{\text{ref}}$ is given by $D\left(\rho||\rho^{\text{ref}}\right)=-\text{Tr}\left[\rho\ln\rho^{\text{ref}}\right]-S\left(\rho\right)$,
where $S\left(\rho\right)=-\text{Tr}\left[\rho\ln\rho\right]$ is
the (von Neumann) entropy \cite{Umegaki1954,Vedral2002,Wilde2017}.
We conventionally write the instantaneous Gibbs equilibrium state
with Hamiltonian $H\left(t\right)$ and inverse temperature $\beta_{i}$
as $\rho_{t}^{\text{eq,}i}=e^{-\beta_{i}H\left(t\right)}/Z_{t}^{i}$,
where $Z_{t}^{i}=\text{Tr }e^{-\beta_{i}H\left(t\right)}$ is the
partition function and $i\in\left\{ \text{c},\text{h}\right\} $ denotes
the cold and hot thermal states, respectively. When the reference
state of the divergence is some thermal state $\rho_{t}^{\text{eq,}i}$,
we will call $D\left(\rho_{t}||\rho_{t}^{\text{eq,}i}\right)$ the
thermal divergence.

For the driving strokes, we define the states $\rho_{t}^{\text{qs,}i}$,
with $i\in\left\{ \text{c},\text{h}\right\} $, as the states that
would have been obtained if the driving was performed quasistatically
(without transition among the instantaneous eigenstates) and if the
initial state was the thermal state $\rho_{t_{0}}^{\text{eq,}i}$,
where $t_{0}=0$ ($t_{0}=\tau_{2}$) for the expansion (compression)
stroke. More explicitly, denoting by $\Ket{E_{n}^{t}}$ the instantaneous
energy eigenstates, the two states associated with the end of the
expansion and compression strokes are $\rho_{\tau_{1}}^{\text{qs,c}}=\sum_{n}p_{n}^{\text{eq,c},0}\Ket{E_{n}^{\tau_{1}}}\Bra{E_{n}^{\tau_{1}}}$
and $\rho_{\tau_{3}}^{\text{qs,h}}=\sum_{n}p_{n}^{\text{eq,h},\tau_{1}}\Ket{E_{n}^{0}}\Bra{E_{n}^{0}}$,
where $p_{n}^{\text{eq,}i,t}$, with $i\in\left\{ \text{c},\text{h}\right\} $,
are the Boltzmann weights calculated with inverse temperature $\beta_{i}$
and Hamiltonian $H\left(t\right)$. When the reference state of the
divergence is the quasistatically evolved state $\rho_{t}^{\text{qs,}i}$,
we call $D\left(\rho_{t}||\rho_{t}^{\text{qs,}i}\right)$ the quasistatic
divergence.

The efficiency of the quantum heat engine is given by the ratio of
the net extracted work over the heat absorbed from the hot source,
i.e., $\eta=-\left\langle W_{\text{net}}\right\rangle /\left\langle Q_{\text{h}}\right\rangle $,
with $\left\langle W_{\text{net}}\right\rangle =\left\langle W_{1}\right\rangle +\left\langle W_{3}\right\rangle <0$
and $\left\langle Q_{\text{h}}\right\rangle >0$. The efficiency of
the QOHE can be related to the total entropy produced ($\left\langle \Sigma_{\text{total}}\right\rangle $)
in a cycle through~\cite{Peterson2018}\textbf{ }(see \ref{subsec:The-thermal-efficiency})
\begin{equation}
\eta=\eta_{\text{Carnot}}-\frac{\left\langle \Sigma_{\text{total}}\right\rangle }{\beta_{\text{c}}\left\langle Q_{\text{h}}\right\rangle },\label{eq:efficiency carnot}
\end{equation}
where $\eta_{\text{Carnot}}=1-\beta_{\text{h}}/\beta_{\text{c}}$.

The entropy produced during the incomplete thermalization with a Markovian
heat reservoir (with constant Hamiltonian) is~\cite{Deffner2011}
$\left\langle \Sigma\right\rangle =D\left(\rho_{0}||\rho^{\text{eq}}\right)-D\left(\rho_{\tau}||\rho^{\text{eq}}\right)\geq0$,
where $\rho_{0}$ and $\rho_{\tau}$ are the initial and final states
of the thermalization process and $\rho^{\text{eq}}$ is the Gibbs
state. Applying these results to the engine cycle one finds (see \ref{subsec:The-thermal-efficiency})
\begin{equation}
\left\langle \Sigma_{\text{total}}\right\rangle =D\left(\rho_{\tau_{1}}||\rho_{\tau_{1}}^{\text{eq,h}}\right)-D\left(\rho_{\tau_{2}}||\rho_{\tau_{2}}^{\text{eq,h}}\right)+D\left(\rho_{\tau_{3}}||\rho_{\tau_{3}}^{\text{eq,c}}\right),\label{eq:entropy production total}
\end{equation}
which relates the total entropy of the cycle to the thermal divergences
between the key-working-substance states. The quantity $\mathcal{L}_{\text{therm}}=\left\langle \Sigma_{\text{total}}\right\rangle /\beta_{\text{c}}\left\langle Q_{\text{h}}\right\rangle $
has been called the efficiency lag~\cite{Peterson2018} (see also
Ref.~\cite{Campisi2016}) since it quantifies the departure of the
engine efficiency to the Carnot efficiency. In this paper, we will
refer to $\mathcal{\mathcal{L}_{\text{therm}}}$ as the thermal efficiency
lag in order to differentiate it from the other efficiency lag discussed
below.

The total entropy production in Eq.~(\ref{eq:entropy production total})
encompasses both the finite-time effects of the driven and thermalization
dynamics of the engine cycle and directly accounts for the amount
of irreversibility in the engine cycle. Through Eq.~(\ref{eq:efficiency carnot}),
since $\left\langle \Sigma_{\text{total}}\right\rangle $ is nonnegative,
the quantum engine efficiency is upper bounded by the Carnot efficiency.

Let $\rho^{\text{ref}}=\sum_{n}\lambda_{n}\Pi_{n}$ denote the spectral
decomposition of some reference state used to compute the divergence,
where $\left\{ \lambda_{n}\right\} $ are the eigenvalues and $\left\{ \Pi_{n}\right\} $
are the eigenprojectors of $\rho^{\text{ref}}$. The divergence $D\left(\rho||\rho^{\text{ref}}\right)$
can be shown to be decomposed as~\cite{Santos2017,Francica2017,Janzing2006,Lostaglio2015}
\begin{equation}
D\left(\rho||\rho^{\text{ref}}\right)=D\left(\varepsilon\left(\rho\right)||\rho^{\text{ref}}\right)+\mathcal{C}\left(\rho\right),\label{eq:decomposition}
\end{equation}
where $\varepsilon\left(\cdot\right)=\sum_{n}\Pi_{n}\left(\cdot\right)\Pi_{n}$
is the full dephasing map and $\mathcal{C}\left(\rho\right)=S\left(\varepsilon\left(\rho\right)\right)-S\left(\rho\right)$
is the relative entropy of coherence (in the reference state basis)~\cite{Aberg2006,Baumgratz2014,Streltsov2017,Winter2016,Singh2017}.

From now on, we conveniently assume that the energy basis of the instantaneous
Hamiltonian $H\left(t\right)$ of the working substance is the relevant
basis where the full dephasing $\varepsilon\left(\rho_{t}\right)$
and the relative entropy of coherence $\mathcal{C}\left(\rho_{t}\right)$
are computed. Applying the decomposition in Eq.~(\ref{eq:decomposition})
to Eq.~(\ref{eq:entropy production total}), the entropy production
can be written as 
\begin{equation}
\left\langle \Sigma_{\text{total}}\right\rangle =\left\langle \Sigma_{\text{total}}^{\text{pop}}\right\rangle +\left\langle \Sigma_{\text{total}}^{\text{coh}}\right\rangle ,\label{eq:entropy production decomposition}
\end{equation}
where 
\begin{align}
\left\langle \Sigma_{\text{total}}^{\text{pop}}\right\rangle = & [D\left(\varepsilon\left(\rho_{\tau_{1}}\right)||\rho_{\tau_{1}}^{\text{eq,h}}\right)-D\left(\varepsilon\left(\rho_{\tau_{2}}\right)||\rho_{\tau_{2}}^{\text{eq,h}}\right)\nonumber \\
 & +D\left(\varepsilon\left(\rho_{\tau_{3}}\right)||\rho_{\tau_{3}}^{\text{eq,c}}\right)]
\end{align}
and 
\begin{equation}
\left\langle \Sigma_{\text{total}}^{\text{coh}}\right\rangle =\mathcal{C}\left(\rho_{\tau_{1}}\right)-\mathcal{C}\left(\rho_{\tau_{2}}\right)+\mathcal{C}\left(\rho_{\tau_{3}}\right)\label{eq:coherent efficiency lag}
\end{equation}
quantify the contribution of the populations ($\left\langle \Sigma_{\text{total}}^{\text{pop}}\right\rangle $)
and coherences ($\left\langle \Sigma_{\text{total}}^{\text{coh}}\right\rangle $)
of the key-working-substance states to the total entropy production,
respectively. These terms show explicitly how the coherence of the
key-working-substance states of the QOHE contributes to the engine
irreversibility and, thus, to the engine efficiency by means of Eq.~(\ref{eq:efficiency carnot}).
Equation~(\ref{eq:efficiency carnot}) is obtained assuming a closed
cycle, without assuming any particular model for the quantum working
substance. Before we discuss the effects of coherence and the dynamical
interference in the engine performance, we present a particular relation
for a QOHE fueled by a single-qubit working substance.

When all strokes take infinite time, i.e., the engine operates in
the quasistatic regime, it does not produce any quantum friction and,
thus, achieves its maximum efficiency, producing minimum entropy along
the cycle. For a QOHE, the maximum efficiency is given by the quantum
Otto efficiency $\eta_{\text{Otto}}=1-\omega_{0}/\omega_{\tau_{1}}$~\cite{Kosloff2017}.

For a single-qubit working substance, we obtained the expression (see
\ref{sec:The-quasistatic-divergences} and \ref{sec:The-quasistatic-efficiency lag})
\begin{equation}
\eta=\eta_{\text{Otto}}-\frac{\mathcal{F}}{\beta_{\text{c}}\left\langle Q_{\text{h}}\right\rangle },\label{eq:efficiency otto}
\end{equation}
where 
\begin{align}
\mathcal{F}= & D\left(\rho_{\tau_{1}}||\rho_{\tau_{1}}^{\text{qs,c}}\right)\nonumber \\
 & +\frac{\omega_{0}\beta_{\text{c}}}{\omega_{\tau_{1}}\beta_{\text{h}}}\left[D\left(\rho_{\tau_{3}}||\rho_{\tau_{3}}^{\text{qs,h}}\right)-D\left(\rho_{\tau_{2}}||\rho_{\tau_{2}}^{\text{eq,h}}\right)\right]\label{eq:quantum friction lag}
\end{align}
quantifies the quantum friction ($\mathcal{F}\geq0$) in the quantum
Otto heat engine~\cite{Marcela}. It contains two quasistatic divergences
(associated with both driven strokes) and the thermal divergence of
the state incompletely thermalized. In the limit that $\tau_{1}$,
$\tau_{2}$, and $\tau_{3}$ are infinitely large, all divergences
are identically zero. This means that no quantum friction implies
$\mathcal{F}=0$, which implies $\eta=\eta_{\text{Otto}}$. We will
refer to the quantity $\mathcal{L}_{\text{qs}}=\mathcal{F}/\beta_{\text{c}}\left\langle Q_{\text{h}}\right\rangle $
as the quasistatic efficiency lag.

Employing the same reasoning that leads to Eq.~(\ref{eq:decomposition}),
we can split the quantum friction into two contributions 
\begin{equation}
\mathcal{F}=\mathcal{F}^{\text{pop}}+\mathcal{F}^{\text{coh}},\label{eq:quantum friction decomposition}
\end{equation}

where 
\begin{align}
\mathcal{F}^{\text{pop}}= & D\left[\varepsilon\left(\rho_{\tau_{1}}\right)||\rho_{\tau_{1}}^{\text{qs,c}}\right]\nonumber \\
 & +\frac{\omega_{0}\beta_{\text{c}}}{\omega_{\tau_{1}}\beta_{\text{h}}} \left\lbrace D\left[\varepsilon\left(\rho_{\tau_{3}}\right)||\rho_{\tau_{3}}^{\text{qs,h}}\right]-D\left[\varepsilon\left(\rho_{\tau_{2}}\right)||\rho_{\tau_{2}}^{\text{eq,h}}\right] \right\rbrace
\end{align}
and 
\begin{equation}
\mathcal{F}^{\text{coh}}=\mathcal{C}\left(\rho_{\tau_{1}}\right)+\frac{\omega_{0}\beta_{\text{c}}}{\omega_{\tau_{1}}\beta_{\text{h}}}\left[\mathcal{C}\left(\rho_{\tau_{3}}\right)-\mathcal{C}\left(\rho_{\tau_{2}}\right)\right]\label{eq:quantum friction coherent part}
\end{equation}
quantify the contribution of the populations ($\mathcal{F}^{\text{pop}}$)
and coherences ($\mathcal{F}^{\text{coh}}$) of the key-working-substance
states, respectively.

The entropy production and quantum friction in Eqs.~(\ref{eq:entropy production total})
and (\ref{eq:quantum friction lag}) have been decomposed into a population
and a coherent contribution with respect to the relevant instantaneous
energy basis. However, this does not mean that the population part
does not depend on the coherences whatsoever. For a driving that generates
coherence in the instantaneous energy eigenbasis, such as our first
and third driven strokes, the populations of the final state depend
on the way that coherence was generated, i.e., depend on the process.
Therefore, the separation into the population and coherent contributions
is with respect to the populations and coherences of the key-working-substance
states.

Combining Eqs.~(\ref{eq:efficiency carnot}) and (\ref{eq:efficiency otto})
one can find the relation~\cite{Peterson2018,Marcela} 
\begin{equation}
\left\langle \Sigma_{\text{total}}\right\rangle =\beta_{\text{c}}\left\langle Q_{\text{h}}\right\rangle \left[\eta_{\text{Carnot}}-\eta_{\text{Otto}}\right]+\mathcal{F}\label{eq:relation entropy production and quantum friction}
\end{equation}
between the total entropy production and the quantum friction in a
QOHE. The minimum entropy production is obtained when there is no
quantum friction, i.e., when the engine runs in the quasistatic regime,
and it is given by $\left\langle \Sigma_{\text{total}}^{\text{min}}\right\rangle =\beta_{\text{c}}\left\langle Q_{\text{h}}^{\text{qs}}\right\rangle \left[\eta_{\text{Carnot}}-\eta_{\text{Otto}}\right]$,
where $\left\langle Q_{\text{h}}^{\text{qs}}\right\rangle $ is the
heat absorbed in the quasistatic limit of the cycle.

\begin{figure*}[t]
\begin{centering}
\includegraphics[scale=0.7]{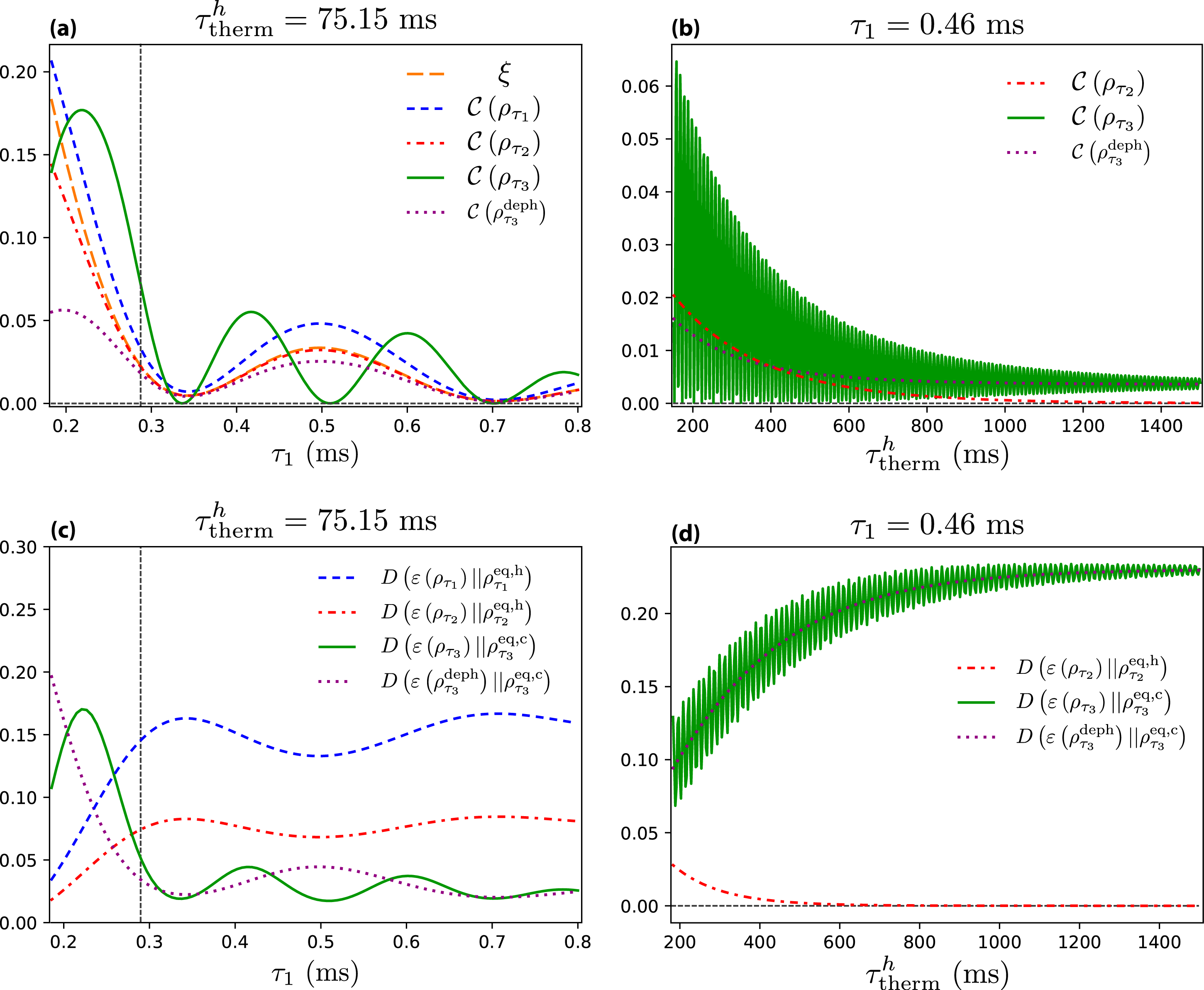} 
\par\end{centering}
\caption{Dynamics of the coherences. \textbf{(a)} The relative entropy of coherences
of the key-working-substance states $\mathcal{C}\left(\rho_{\tau_{1}}\right)$
(blue dashed line), $\mathcal{C}\left(\rho_{\tau_{2}}\right)$ (red
dash-dotted line), and $\mathcal{C}\left(\rho_{\tau_{3}}\right)$
(green solid line) for the original and $\mathcal{C}\left(\rho_{\tau_{3}}^{\text{deph}}\right)$
(purple dotted line) for dephased engine cycles. Additionally, the
energy transition probability $\xi^{\text{exp}}\left(\tau_{1},0\right)$
(orange long-dashed line) is also shown. All quantities are plotted
assuming a fixed thermalization time $\tau_{\text{therm}}^{\text{h}}=75.15\text{ ms}$
and varying the driving time $\tau_{1}$. The engine works as a heat
engine for the parameters on the right of the dashed vertical line.
\textbf{(b)} The relative entropy of coherence of the relevant key-working-substance
states as a function of the thermalization time $\tau_{\text{therm}}^{\text{h}}$
for fixed driving time $\tau_{1}=0.46\text{ ms}$.\textcolor{red}{{}
}\textbf{(c)} The population component of the thermal divergence of
the key-working-substance states, $D\left(\varepsilon\left(\rho_{\tau_{1}}\right)||\rho_{\tau_{1}}^{\text{eq,h}}\right)$
(blue dashed line), $D\left(\varepsilon\left(\rho_{\tau_{2}}\right)||\rho_{\tau_{2}}^{\text{eq,h}}\right)$
(red dash-dotted line), and $D\left(\varepsilon\left(\rho_{\tau_{3}}\right)||\rho_{\tau_{3}}^{\text{eq,c}}\right)$
(green solid line) for the original and $D\left(\varepsilon\left(\rho_{\tau_{3}}^{\text{deph}}\right)||\rho_{\tau_{3}}^{\text{eq,c}}\right)$
(purple dotted line) for the dephased engine cycles. All quantities
are plotted assuming a fixed thermalization time $\tau_{\text{therm}}^{\text{h}}=75.15\text{ ms}$
and varying the driving time $\tau_{1}$. \textbf{(d)} The population
component of the thermal divergence of the key-working-substance states
as a function of the thermalization time $\tau_{\text{therm}}^{\text{h}}$
for fixed driving time $\tau_{1}=0.46\text{ ms}$. \label{fig:Coherences.}
The map $\varepsilon$ is the full dephasing operation as explained
in Eq. (\ref{eq:decomposition}). All entropic quantities are computed
in the natural unit of information (nat).}
\end{figure*}

The engine average power output per cycle is given by $P_{\text{tot}}=-\left\langle W_{\text{net}}\right\rangle /\tau_{\text{cycle}}$,
where $\tau_{\text{cycle}}=\tau_{4}=\tau_{1}+\tau_{\text{therm}}^{\text{h}}+\tau_{1}+\tau_{\text{therm}}^{\text{c}}$
is the cycle time duration. The relation between the efficiency and
power is given by $P_{\text{tot}}=\eta\left\langle Q_{\text{h}}\right\rangle /\tau_{\text{cycle}}$
. With this expression and Eqs.~(\ref{eq:efficiency carnot}) and
(\ref{eq:efficiency otto}), the extractable power can be written
in terms of the entropy production or quantum friction as 
\begin{equation}
P_{\text{tot}}=\frac{\eta_{\text{Carnot}}\left\langle Q_{\text{h}}\right\rangle }{\tau_{\text{cycle}}}-\frac{\left\langle \Sigma_{\text{total}}\right\rangle }{\beta_{\text{c}}\tau_{\text{cycle}}}
\end{equation}
and 
\begin{equation}
P_{\text{tot}}=\frac{\eta_{\text{Otto}}\left\langle Q_{\text{h}}\right\rangle }{\tau_{\text{cycle}}}-\frac{\mathcal{F}}{\beta_{\text{c}}\tau_{\text{cycle}}},
\end{equation}
respectively. Then, using the decompositions in Eqs.~(\ref{eq:entropy production decomposition})
and (\ref{eq:quantum friction decomposition}), the power can be written
in terms of the relative entropy of coherence of the key-working-substance
states.

The expressions discussed above for the finite-time dynamics of the
engine explicitly show how the energy coherence of the key-working-substance
states contributes to the engine performance and irreversibility. At
this point it is important to clarify that, for an engine which generates
coherence in the energy basis during the first stroke, a complete
thermalization with the hot heat reservoir would imply $D\left(\rho_{\tau_{2}}||\rho_{\tau_{2}}^{\text{eq,h}}\right)=0$
and $\mathcal{C}\left(\rho_{\tau_{2}}\right)=0$ identically. Hence,
the presence of coherence would always contribute to an increase in
irreversibility (both entropy production and quantum friction) and
a decrease in the engine performance, when compared with an engine
where the driven strokes do not produce coherence.

On the other hand, if the thermalization is incomplete, the contribution
of the incompletely thermalized state $\rho_{\tau_{2}}$ (after the
second stroke) is manifested through the term $D\left(\rho_{\tau_{2}}||\rho_{\tau_{2}}^{\text{eq,h}}\right)$.
Such a thermal divergence contributes to decrease both the entropy
production and quantum friction as can verified in Eqs.~(\ref{eq:entropy production total})
and (\ref{eq:quantum friction lag}). This means that the residual
coherence after the incomplete thermalization (second stroke), quantified
by $\mathcal{C}\left(\rho_{\tau_{2}}\right)$ in Eqs.~(\ref{eq:quantum friction lag})
and (\ref{eq:quantum friction coherent part}), helps decrease the
irreversibility and, consequently, increase the engine performance
in comparison to an engine that generates coherence and has complete
thermalization. In the next section, we will show that not only does the
residual coherence directly decrease the entropy production and quantum
friction but it also gives rise to a dynamical interference effect that
also helps decrease the irreversibility of the engine.

\subsection{Numerical analysis\label{subsec:Numerical-Analysis}}

Let us now investigate the effects of the coherence in the key-working-substance
states during the cycle, and in particular the residual coherence,
employing numerical simulations. In this analysis, we consider energy
scales compatible with quantum thermodynamics experiments in nuclear
magnetic resonance setups~\cite{Peterson2018,Batalh=00003D0000E3o2014,Batalh=00003D0000E3o2015,Camati2016,Micadei2017}.
The initial and final frequency gaps of the expansion stroke will
be chosen as $\omega_{0}/2\pi=2.0\text{ kHz}$ and $\omega_{\tau_{1}}/2\pi=3.6\text{ kHz}$,
respectively. The chosen temperatures are such that the thermal energy
scale of the cold (hot) heat reservoir is half (double) the energy
gap of the working substance at the time of the interaction with the
heat source. More precisely, the cold and hot inverse temperatures
will be chosen as $\beta_{\text{c}}=2/\left(\hbar\omega_{0}\right)$
and $\beta_{\text{h}}=1/\left(2\hbar\omega_{\tau_{1}}\right)$, respectively.

We assume that a complete thermalization with the cold environment
is approximately achieved at a finite time satisfying the condition
$\tau_{\text{therm}}^{\text{c}}\gg\tau_{\text{relaxation}}^{\text{c}}$.
We also consider that the interaction time with the hot reservoir
is smaller than or of the same order as the relaxation time of the hot
environment $\tau_{\text{therm}}^{\text{h}}\apprle\tau_{\text{relaxation}}^{\text{h}}$,
which results in an incomplete thermalization with the hot environment
in the engine cycle. The strength of the interaction of the working
substance with the hot and cold heat reservoirs, namely, the vacuum
decay rate, will be assumed as $\gamma_{0}^{\text{h}}=\gamma_{0}^{\text{c}}=1\text{Hz}$,
which implies the relaxation time $\tau_{\text{relaxation}}^{\text{h}}=244.92\text{ ms}$
and $\tau_{\text{relaxation}}^{\text{c}}=761.59\text{ ms}$. In order
to reach approximately a complete thermalization at the end of the
fourth stroke, we considered the cold thermalization time as $\tau_{\text{therm}}^{\text{c}}\approx6.56\times\tau_{\text{relaxation}}^{\text{c}}$.
In this case, the working substance approximately returns to the cold
Gibbs state. The trace distance between the final state of the fourth
stroke and the cold Gibbs state is approximately $\frac{1}{2}\text{Tr}\left|\rho_{\tau_{4}}-\rho_{0}^{\text{eq,c}}\right|\simeq10^{-2}$.

Figures.~\ref{fig:Coherences.}\textbf{(a)} and \ref{fig:Coherences.}\textbf{(b)}
display the relative entropies of coherence as a function of the driving
time $\tau_{1}$ and the thermalization time $\tau_{\text{therm}}^{\text{h}}$,
respectively. In Fig.~\ref{fig:Coherences.}\textbf{(a)} one can
see that the relative entropy of coherence at the end of the first
and second strokes are qualitatively similar. The residual coherence
$\mathcal{C}\left(\rho_{\tau_{2}}\right)$ is smaller than $\mathcal{C}\left(\rho_{\tau_{1}}\right)$
due to a partial decrease of $\mathcal{C}\left(\rho_{\tau_{1}}\right)$
set by the incomplete thermalization.

Recall that the quantities for the dephased engine cycle are labeled
by the superscript ``deph'' {[}see caption of Fig.~\ref{fig:Generic-engine-cycle.}\textbf{(b)}{]}.
The coherence at the end of the third stroke of the dephased engine
cycle $\mathcal{C}\left(\rho_{\tau_{3}}^{\text{deph}}\right)$ behaves
qualitatively as $\mathcal{C}\left(\rho_{\tau_{1}}\right)$ and $\mathcal{C}\left(\rho_{\tau_{2}}\right)$,
even being smaller than $\mathcal{C}\left(\rho_{\tau_{2}}\right)$.
Note how the behavior of the coherence at the end of the third stroke
in the original (not dephased) engine cycle $\mathcal{C}\left(\rho_{\tau_{3}}\right)$
is different. This qualitative difference comes from the interference
between the residual coherence and the coherence generated at the
third stroke [see green and purple dotted curves in Figs.~\ref{fig:Coherences.}\textbf{(a)}
and \ref{fig:Coherences.}\textbf{(b)}].

Figures \ref{fig:Coherences.}\textbf{(c)} and \ref{fig:Coherences.}\textbf{(d)}
show the population component of the thermal divergence as a function
of the driving time $\tau_{1}$ and the thermalization time $\tau_{\text{therm}}^{\text{h}}$,
respectively. Again, in \ref{fig:Coherences.}\textbf{(c) }one can
note that the curves for the population component of the thermal divergence
at the end of the first and second strokes are similar. Furthermore,
the behaviors of the population components of the thermal divergences
at the end of the third stroke of the original {[}$D\left(\varepsilon\left(\rho_{\tau_{3}}\right)||\rho_{\tau_{3}}^{\text{eq,c}}\right)${]}
and the dephased {[}$D\left(\varepsilon\left(\rho_{\tau_{3}}^{\mbox{deph}}\right)||\rho_{\tau_{3}}^{\text{eq,c}}\right)${]}
engines are qualitatively different. Such a difference also comes
from the dynamical interference effect, highlighting that the populations
of the key-working-substance states are not independent from the generated
coherence during the strokes [see green and dotted-purple curves in
Figs. \ref{fig:Coherences.}\textbf{(c)} and \ref{fig:Coherences.}\textbf{(d)}].

The amount of coherence as measured by $\mathcal{C}\left(\rho_{\tau_{2}}\right)$
decays exponentially with the thermalization time $\tau_{\text{therm}}^{\text{h}}$
[see Fig.~\ref{fig:Coherences.}\textbf{(b)}]. On the other hand, the
coherence $\mathcal{C}\left(\rho_{\tau_{3}}\right)$ oscillates quickly
due to the interference of the residual coherence $\mathcal{C}\left(\rho_{\tau_{2}}\right)$
and the coherence generated by the driven dynamics in the compression
stroke. As the thermalization time increases, the oscillating amplitudes
of $\mathcal{C}\left(\rho_{\tau_{3}}\right)$ become less pronounced,
going asymptotically to zero, in which case the coherence $\mathcal{C}\left(\rho_{\tau_{3}}\right)$
approaches $\mathcal{C}\left(\rho_{\tau_{3}}^{\text{deph}}\right)$
because the coherence $\mathcal{C}\left(\rho_{\tau_{1}}\right)$ is
increasingly erased. In the expressions for the efficiency and power
output, the term $\mathcal{C}\left(\rho_{\tau_{2}}\right)-\mathcal{C}\left(\rho_{\tau_{3}}\right)$
explicitly appears, suggesting that whenever $\mathcal{C}\left(\rho_{\tau_{2}}\right)\geq\mathcal{C}\left(\rho_{\tau_{3}}\right)$
efficiency and power output can be enhanced, compared to the dephased
engine performance. From Fig.~\ref{fig:Coherences.}\textbf{(b)}
we can see that the rapid oscillations make this inequality be satisfied
for very narrow time intervals. Such a behavior will be present in
the efficiency and power output as will be seen shortly.

\begin{figure*}[t]
\begin{centering}
\includegraphics[scale=0.72]{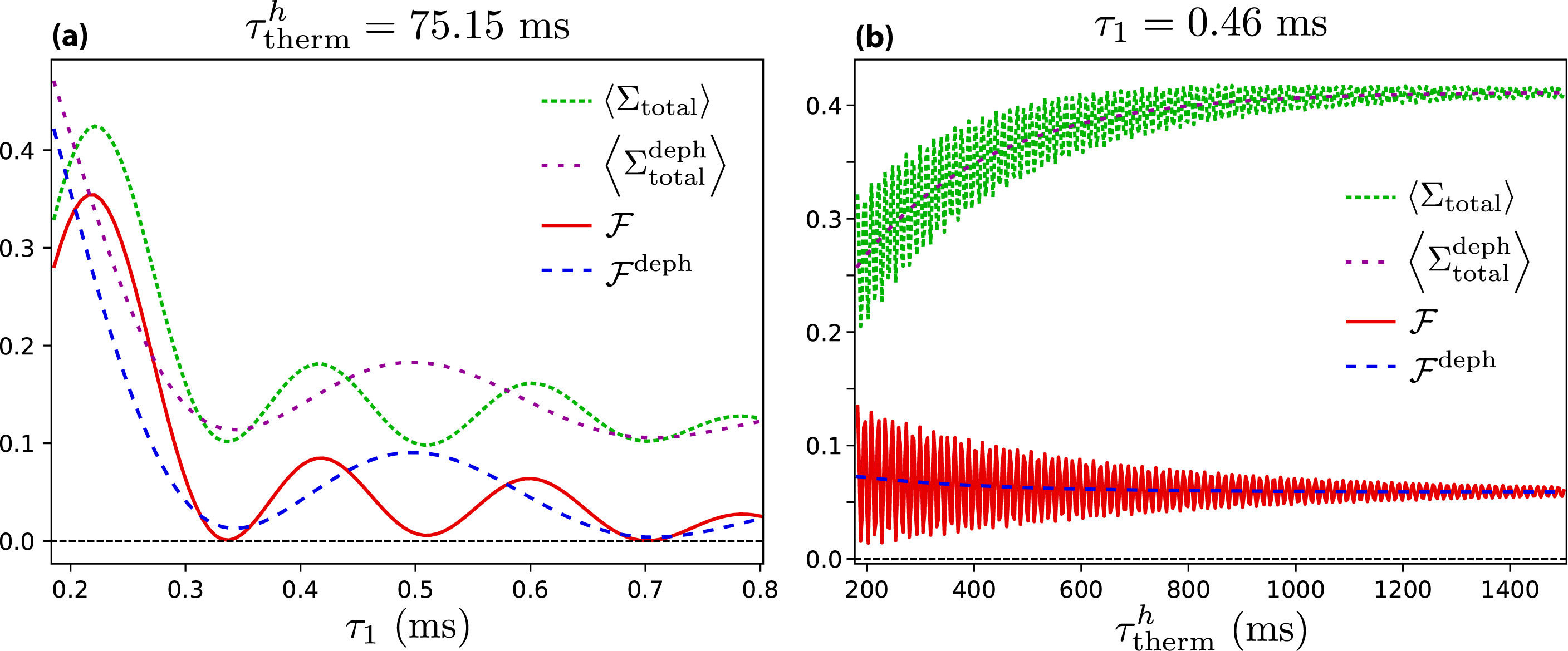} 
\par\end{centering}
\caption{\textcolor{black}{Entropy production and quantum friction as a function
of the driving time $\tau_{1}$}\textbf{\textcolor{black}{{} (a)}}\textcolor{black}{{}
and thermalization time $\tau_{\text{therm}}^{\text{h}}$}\textbf{\textcolor{black}{{}
(b)}}\textcolor{black}{. Total entropy production of the engine with
and without dynamical interference is $\left\langle \Sigma_{\text{total}}\right\rangle $
and $\left\langle \Sigma_{\text{total}}^{\text{deph}}\right\rangle $,
respectively. Similarly, the quantum friction of the engine with and
without dynamical interference is $\mathcal{F}$ and $\mathcal{F}^{\text{deph}}$,
respectively. All entropic quantities are computed in the natural
unit of information (nat).\label{fig:efficiency lags}}}
\end{figure*}

In Fig. \ref{fig:Coherences.}\textbf{(d)}, one can observe that $D\left(\varepsilon\left(\rho_{\tau_{2}}\right)||\rho_{\tau_{2}}^{\text{eq,h}}\right)$
approaches zero as the thermalization time increases, as expected.
On the other hand, the oscillatory profile in $D\left(\varepsilon\left(\rho_{\tau_{3}}\right)||\rho_{\tau_{3}}^{\text{eq,c}}\right)$
(green solid line) in comparison to $D\left(\varepsilon\left(\rho_{\tau_{3}}^{\text{deph}}\right)||\rho_{\tau_{3}}^{\text{eq,c}}\right)$
(purple dotted line) is solely due to the interference of the residual
coherence and the coherence generated during the third stroke. It
is interesting to note that, in the present scenario, the relative
entropies of coherence and the population components of the thermal
divergences can not be varied in an independent way varying the stroke
parameters. For instance, in Fig.~\ref{fig:Coherences.}\textbf{(a)}
and \ref{fig:Coherences.}\textbf{(c)}, $\mathcal{C}\left(\rho_{\tau_{3}}\right)$
and $D\left(\varepsilon\left(\rho_{\tau_{3}}^{\text{deph}}\right)||\rho_{\tau_{3}}^{\text{eq,c}}\right)$
oscillate roughly in phase while $\mathcal{C}\left(\rho_{\tau_{2}}\right)$
and $D\left(\varepsilon\left(\rho_{\tau_{2}}\right)||\rho_{\tau_{2}}^{\text{eq,h}}\right)$
oscillate roughly out of phase. This aspect highlights the interrelation
between the coherences and populations.

Figures~\ref{fig:efficiency lags}\textbf{(a)} and \ref{fig:efficiency lags}\textbf{(b)
}display the total entropy production and quantum friction of the
original (with dynamical interference) and the dephased (without dynamical
interference) engines as a function of the driving time $\tau_{1}$
and the thermalization time $\tau_{\text{therm}}^{\text{h}}$, respectively.
The entropy production $\left\langle \Sigma_{\text{total}}\right\rangle $
and quantum friction $\mathcal{F}$ differ by a term containing the
heat absorbed by the working substance from the hot heat reservoir
and the difference between the Carnot and Otto efficiencies {[}see
Eq.~(\ref{eq:relation entropy production and quantum friction}){]}.
The difference of the efficiency lags, on the other hand, is constant
with respect to $\tau_{1}$ or $\tau_{\text{therm}}^{\text{h}}$,
$\mathcal{L}_{\text{therm}}-\mathcal{L}_{\text{qs}}=\eta_{\text{Carnot}}-\eta_{\text{Otto}}$.

Comparing the entropy production and quantum friction of either the
original or the dephased engine to each other, Fig.~\ref{fig:efficiency lags}\textbf{(a)}
shows that their qualitative behaviors are the same with respect to
the driving time $\tau_{1}$. Moreover, note that the quantities for
the original and dephased engine change in different timescales. The
original engine quantities contain the effect of the dynamical interference
and behave similarly to the relative entropy of coherence $C\left(\rho_{\tau_{3}}\right)$
and population divergence $D\left[ \varepsilon\left(\rho_{\tau_{3}}\right)||\rho_{\tau_{3}}^{\text{eq,c}}\right]$
[see Figs.~\ref{fig:Coherences.}\textbf{(a)} and \ref{fig:Coherences.}\textbf{(c)}].

The oscillation timescale of entropy production and quantum friction
with respect to the thermalization time $\tau_{\text{therm}}^{\text{h}}$
shown in Fig.~\ref{fig:efficiency lags}\textbf{(b)} also agree with
the relative entropy of coherence and population divergence {[}compare
with Figs.~\ref{fig:Coherences.}\textbf{(b)} and \ref{fig:Coherences.}\textbf{(d)}{]}.
However, entropy production increases while the quantum friction decreases
the more the working substance reaches the thermal state. The closer
a system reaches to the thermal state the larger the entropy produced~\cite{Deffner2011},
so the increase in Fig.~\ref{fig:efficiency lags}\textbf{(b)} is
expected. Increasing the thermalization time decreases the coherence
generated in the third stroke {[}see Fig.~\ref{fig:Coherences.}\textbf{(b)}{]}.
The decrease in quantum friction is associated with the decrease in
the total amount of coherence generated in the engine, showing the
deep connection of quantum friction and energy coherence.

Our results generalize and qualitatively explain some previous findings
in the literature. For instance, in Refs.~\cite{Rezek2010,Feldmann2006}
noise has been used to improve the quantum engine efficiency. Here
we observe that, if the noise is such that it decreases the contribution
of either $\mathcal{C}\left(\rho_{\tau_{1}}\right)$ or $\mathcal{C}\left(\rho_{\tau_{3}}\right)$,
the efficiency can be enhanced. Moreover, quite a few papers have
considered the so-called energy entropy of the working substance
as a measure of quantum friction~\cite{Feldmann2004,Rezek2010,Feldmann2012,Zagoskin2012}.
Although not directly connected in these works, the difference between
what these authors called energy entropy and the von Neumann entropy
is nothing but the relative entropy of coherence. The present paper
elucidates how this energy entropy was related to quantum friction
in Refs. \cite{Feldmann2004,Rezek2010,Feldmann2012,Zagoskin2012},
namely, through the relative entropy of coherence in Eqs.~(\ref{eq:quantum friction lag}),
(\ref{eq:quantum friction decomposition}), and (\ref{eq:quantum friction coherent part}).
In Refs.~\cite{Feldmann2006,Rezek2006,Feldmann2012,Thomas2014,Correa2015},
quantum friction has been somewhat related to the presence of coherence.
Reference~\cite{Feldmann2012}, for instance, employed the $l_{1}$ norm
of coherence to quantify quantum friction. Our results complement
these findings by providing a concrete relation that elucidates how
energy coherence is linked to quantum friction by means of the quasistatic
and thermal divergences.

\begin{figure*}
\begin{centering}
\includegraphics[scale=0.68]{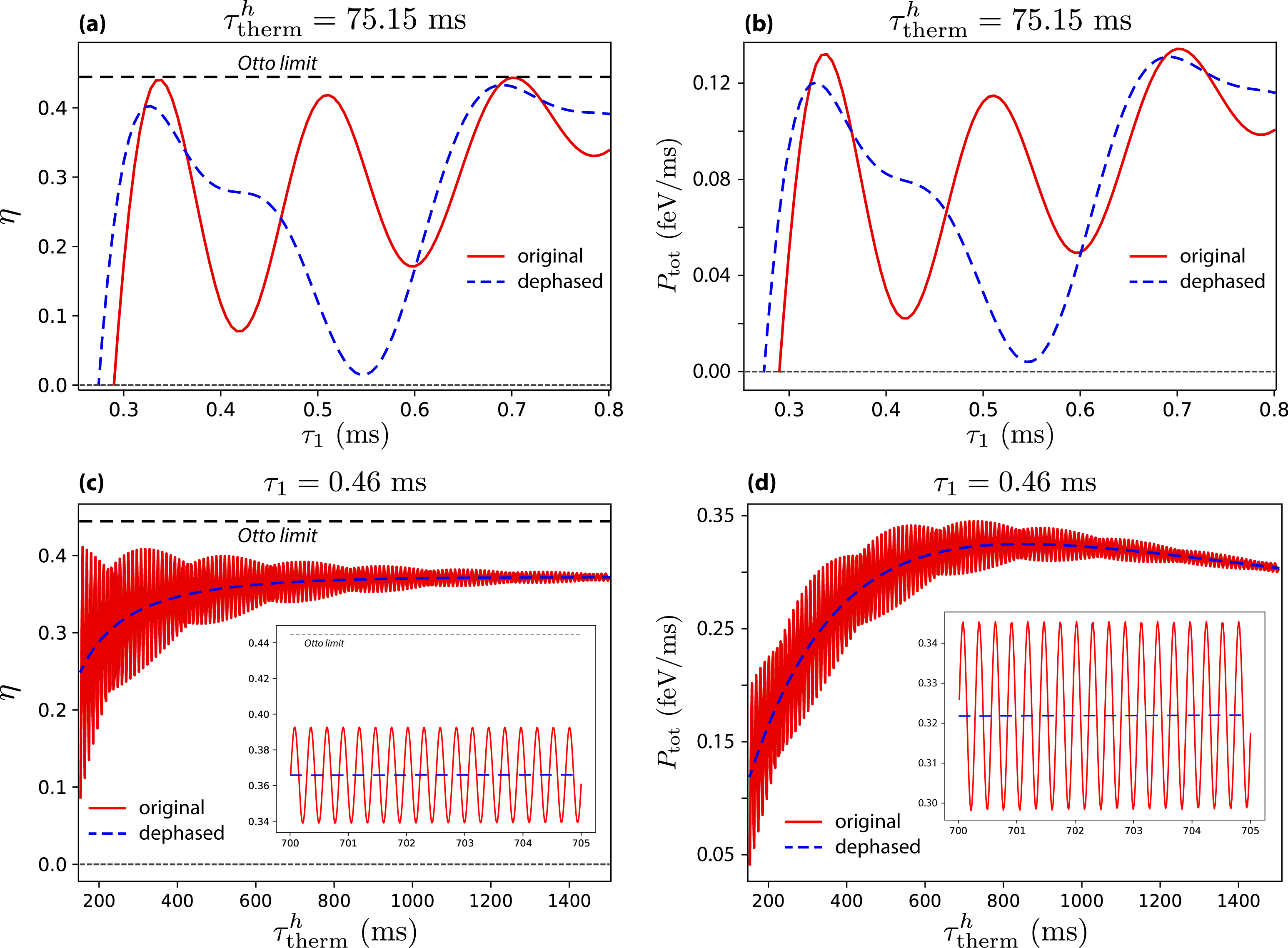} 
\par\end{centering}
\caption{Efficiency and power output. \textbf{(a)} and \textbf{(b)} The engine
efficiency and power output as a function of the driving time $\tau_{1}$
for the fixed thermalization time $\tau_{\text{therm}}^{\text{h}}=75.15\text{ ms}$,
respectively. \textbf{(c)} and \textbf{(d)} The engine efficiency
and power output as a function of the thermalization time $\tau_{\text{therm}}^{\text{h}}$
for the fixed driving time $\tau_{1}=0.46\text{ ms}$, respectively.
In all figures, the efficiency and power output of the original and
dephased engine cycle are depicted in red solid line and blue dashed
line, respectively. \label{fig:Thermodynamic-Quantities.}}
\end{figure*}

We have seen some effects of the interference of coherence in the
previous plots and discussions. We further analyze this phenomenon
by showing exactly how it contributes energetically to the thermodynamic
quantities in the quantum cycle. The following relations have all
been obtained assuming a single qubit as working substance, as described
in Sec.~\ref{sec:The-Engine-Cycle}.

Let us denote by $E_{n}^{t}$ and $\Ket{E_{n}^{t}}$ the instantaneous
eigenenergies and eigenstates of the engine Hamiltonian, respectively,
where the index $n=0$ ($n=1$) stands for the ground (excited) state.
The energy transition probability in the first stroke is given by
\begin{equation}
p_{\tau_{1},0}^{\text{exp}}\left(1|0\right)=\left|\Bra{E_{1}^{\tau_{1}}}U_{\tau_{1},0}\Ket{E_{0}^{0}}\right|^{2}=\xi\left(\tau_{1},0\right).\label{eq:energy transition probability}
\end{equation}
Evaluating the first-stroke work and the second-stroke heat one obtains
$\left\langle W_{1}\right\rangle =\frac{1}{2}\left\{ \hbar\omega_{0}-\hbar\omega_{\tau_{1}}\left[1-2\xi\left(\tau_{1},0\right)\right]\right\} g_{\text{c}}$
and $\left\langle Q_{\text{h}}\right\rangle =\frac{1}{2}\hbar\omega_{\tau_{1}}\left\{ \left[1-2\xi\left(\tau_{1},0\right)\right]g_{\text{c}}+r_{z}\left(\tau_{2}\right)\right\} $,
respectively, where $g_{\text{c}}=\tanh\left(\frac{1}{2}\beta_{\text{c}}\hbar\omega_{0}\right)$
and $r_{z}\left(\tau_{2}\right)=\text{Tr}\left[\sigma_{z}\rho_{\tau_{2}}\right]$
is the $z$ component of the qubit Bloch vector associated with $\rho_{\tau_{2}}$.
In particular, for a complete thermalization, this Bloch vector component
will be $r_{z}\left(\tau_{2}\rightarrow\infty\right)=g_{\text{h}}=\tanh\left(\frac{1}{2}\beta_{\text{h}}\hbar\omega_{\tau_{1}}\right)$.

The third-stroke work can be evaluated as 
\begin{equation}
\left\langle W_{3}\right\rangle =\frac{1}{2}\left\{ \hbar\omega_{0}\left[1-2\zeta\left(\tau_{3},\tau_{2}\right)\right]-\hbar\omega_{\tau_{1}}\right\} r_{z}\left(\tau_{2}\right)+\mathcal{E}_{\text{inter}},\label{eq:work stroke 3}
\end{equation}
where $\zeta\left(\tau_{3},\tau_{2}\right)=p_{\tau_{3},\tau_{2}}^{\text{com}}\left(1|0\right)=\left|a_{10}^{\text{com}}\left(\tau_{3},\tau_{2}\right)\right|^{2}$
is the third-stroke energy transition probability, $a_{mn}^{\text{com}}\left(\tau_{3},\tau_{2}\right)=\Bra{E_{m}^{\tau_{3}}}V_{\tau_{3},\tau_{2}}\Ket{E_{n}^{\tau_{2}}}$
is the energy probability amplitude, and 
\begin{align}
\mathcal{E}_{\text{inter}}= & 2\sum_{n}E_{n}^{0}e^{-\gamma^{\text{h}}\tau_{\text{therm}}^{\text{h}}/2}\nonumber \\
 & \times\text{Re}\left\{ \rho_{10}\left(\tau_{1}\right)e^{i\tau_{\text{therm}}^{\text{h}}\omega_{\tau_{1}}}a_{n1}^{\text{com}}a_{n0}^{\text{com}*}\right\} \label{eq:energy trans}
\end{align}
is the energy contribution due to the interference between the residual
coherence and the coherence generated in the third stroke (see \ref{sec:Transferred-coherence-energy-contribution}).
In Eq.~(\ref{eq:energy trans}), $\gamma^{\text{h}}=\gamma_{0}^{\text{h}}\left(2N_{\text{BE}}^{\text{h}}+1\right)$
is the total decay rate of the qubit after the interaction with the
hot heat reservoir, and $\rho_{10}\left(\tau_{1}\right)=\Bra{E_{1}^{\tau_{1}}}\rho_{\tau_{1}}\Ket{E_{0}^{\tau_{1}}}$
is one of the coherence elements of the qubit state in the instantaneous
energy basis at the beginning of the second stroke.

The contribution $\mathcal{E}_{\text{inter}}$ in Eq.~(\ref{eq:work stroke 3})
comes exclusively from the residual coherence at the end of the second
stroke, i.e., the coherence that was not completely erased by incomplete
thermalization. If the thermalization was complete ($\tau_{\text{therm}}^{\text{h}}\rightarrow\infty$)
then $\mathcal{E}_{\text{inter}}=0$. Since the third-stroke work
is present in the engine efficiency and power output, it is clear
that $\mathcal{E}_{\text{inter}}$ {[}Eq.~(\ref{eq:energy trans}){]}
changes the engine performance. Next, we focus on how exactly $\mathcal{E}_{\text{trans}}$
changes the relevant quantities.

The internal energies of the original and dephased QOHEs are related
as: $\mathcal{E}_{0}=\mathcal{E}_{0}^{\text{deph}}$, $\mathcal{E}_{\tau_{1}}=\mathcal{E}_{\tau_{1}}^{\text{deph}}$,
$\mathcal{E}_{\tau_{2}}=\mathcal{E}_{\tau_{2}}^{\text{deph}}$, and
$\mathcal{E}_{\tau_{3}}=\mathcal{E}_{\tau_{3}}^{\text{deph}}+\mathcal{E}_{\text{inter}}$,
where $\mathcal{E}_{\text{trans}}$ is given in Eq.~(\ref{eq:energy trans}).
From these relations we can readily obtain the efficiency 
\begin{equation}
\eta=\eta^{\text{deph}}-\frac{\mathcal{E}_{\text{inter}}}{\left\langle Q_{\text{h}}\right\rangle }\label{eq:comparison efficiencies}
\end{equation}
and power 
\begin{equation}
P_{\text{tot}}=P_{\text{tot}}^{\text{deph}}-\frac{\mathcal{E}_{\text{inter}}}{\tau_{\text{cycle}}}
\end{equation}
of the original engine written with respect to the efficiency and
power of the dephased engine. Furthermore, from Eq.~(\ref{eq:comparison efficiencies})
and Eqs.~(\ref{eq:efficiency carnot}) and (\ref{eq:efficiency otto}),
we can also obtain how the entropy production and quantum friction
change due to the residual coherence; they are given by 
\begin{equation}
\left\langle \Sigma_{\text{total}}\right\rangle =\left\langle \Sigma_{\text{total}}^{\text{deph}}\right\rangle +\beta_{\text{c}}\mathcal{E}_{\text{inter}}\label{eq:22}
\end{equation}
and 
\begin{equation}
\mathcal{F}=\mathcal{F}^{\text{deph}}+\beta_{\text{c}}\mathcal{E}_{\text{inter}},\label{eq:23}
\end{equation}
respectively. All these last four equations show how the performance
of the engine is affected by the interference between the residual
coherence and the coherence generated in the third stroke, quantified
by $\mathcal{E}_{\text{inter}}$. Next, we see this effect in our
particular QOHE.

In Fig.~\ref{fig:Thermodynamic-Quantities.}\textbf{(a)} we compare
the efficiency of the original QOHE (red solid line), which contains
the dynamical interference effect, and the dephased QOHE (blue dashed
line), which does not contain the interference effect, as a function
of the driving time $\tau_{1}$ for a fixed incomplete thermalization
time $\tau_{\text{therm}}^{\text{h}}=75.15\text{ ms}$ ($\tau_{\text{therm}}^{\text{h}}\approx0.31\times\tau_{\text{relaxation}}^{\text{h}}$).
Observing Fig~\ref{fig:Thermodynamic-Quantities.}\textbf{(a),} we
can note that the original QOHE may perform better or worse than the
dephased QOHE. In this parameter regime, the dynamical interference
can make the original QOHE perform about $20$ times more efficiently
than the dephased QOHE in the intermediate region of the plot. Furthermore,
even for such a small thermalization time (about 1/3 of the relaxation
time), the efficiency of the original QOHE reaches values very close
to Otto's efficiency. This is a consequence of the small quantum friction
generated by the engine cycle as seen in Fig.~\ref{fig:efficiency lags}\textbf{(a)}.

A similar behavior can be observed in the power output [see Fig.~\ref{fig:Thermodynamic-Quantities.}\textbf{(b)}].
The power output of the original QOHE can also be greater or smaller
than the power of the dephased QOHE. Note that the efficiency and
power of both the original and dephased QOHE oscillate. Since, by
construction, there is no interference effect in the dephased QOHE,
these oscillations are not a manifestation of the residual coherence
alone. They arise from the choice of the driving Hamiltonian.

Now let us consider a fixed driving time $\tau_{1}$. In this case,
the efficiency and power of the original QOHE oscillate as function
of the thermalization time $\tau_{\text{therm}}^{\text{h}}$ as displayed
in Figs.~\ref{fig:Thermodynamic-Quantities.}\textbf{(c)} and \textbf{(d)}.
Comparing them to Figs.~\ref{fig:Coherences.}\textbf{(a)} and \textbf{(b)},
the oscillations in Figs.~\ref{fig:Thermodynamic-Quantities.}\textbf{(c)}
and \textbf{(d)} occur strictly due to interference between the residual
coherence and the coherence generated in the third stroke. The effect
of this interference is damped as the thermalization time $\tau_{\text{therm}}^{\text{h}}$
increases {[}see Eq.~(\ref{eq:energy trans}){]}.

For an engine that generates coherence in the first stroke, the interference
effect may or may not contribute to increase the engine performance
and function as a quantum lubricant. A fine control over the driving
and thermalization times is paramount to make the quantum engine run
in a suitable parameter regime, thus taking full advantage of dynamical
interference effect.

We emphasize that we are not claiming that the presence of coherence
provides an absolute enhancement of the engine performance. In fact,
from Eqs.~(\ref{eq:coherent efficiency lag}) and (\ref{eq:quantum friction coherent part}),
we see quite the opposite for QOHE engines. We claim that, for a QOHE
that already generates coherence, the residual coherence surviving
the incomplete thermalization stroke may be beneficial to reduce entropy
production and quantum friction if compared to an engine which does
not allow the dynamical interference effect.

Quantum lubrication is a method by which the engine efficiency can
be enhanced through the reduction of quantum friction~\cite{Feldmann2006}
controlling the entropy production and irreversibility along the quantum
cycle. The typical method employed in the literature is to perform
shortcuts to adiabaticity by means of counter-adiabatic driving fields~\cite{Campo2018,Torrontegui2013,Campo2014,Funo2017,Zheng2016,Campbell2017,Calzetta2018}.
For an engine that generates coherence, we have seen that the fine
control over driving and thermalization times can be employed to reduce
entropy production and quantum friction through the interference effect,
when compared to the dephased engine. Since this method does not rely
on additional driving fields but on purely controlling the parameters
of the engine cycle, we refer to it as a dynamical quantum lubrication
strategy.

\section{Conclusions}

In this paper, we have analyzed the role of coherence in a quantum
Otto heat engine. By considering finite-time driven operations and
incomplete thermalization with the hot source, we obtained analytical
expressions relating the entropy production and quantum friction to
the coherence in the energy basis of the working substance along the
cycle. Then, we employed the relationship between these quantities
and the engine efficiency to find how both efficiency and power output
become related to energy coherence. We note that the relation between
entropy production and coherence is valid for any working substance.
Also, we assumed a single-qubit working substance to derive the
relation between the quantum friction and energy coherence.

We found that the residual coherence present at the end of the incomplete
thermalization stroke interferes with the coherence generated in the
compression stroke. In particular, this dynamical interference effect
influences the work performed in the compression stroke. In turn,
this affects the engine efficiency and power output. In order to analyze
this effect, we compared the engine performance to the performance
of a similar engine where the residual coherence is completely erased,
ruling out the dynamical interference effect.

We show that the thermodynamic quantities between the engine with
and without dynamical interference differ by a term which precisely
quantifies the interference effect [cf. Eqs.~(\ref{eq:22}) and (\ref{eq:23})].
We employed a numerical analysis with parameters that can be achieved
with current experimental settings~\cite{Peterson2018}. We numerically
show that the interference is clearly manifested in the engine efficiency
and power output. Therefore, the performance of the engine with dynamical
interference can be either better or worse than that of the engine
without interference. In order to make the engine run in the ``constructive
regime'' (where interference improves the performance), one has to
have a fine control over the cycle parameters (driving and thermalization
times). When operating in this regime, the interference effect can
be seen as a dynamical quantum lubricant.

We believe that our paper contributes to unveil the important role
played by coherence in thermodynamics of quantum devices. We hope
that the results presented here encourage new experimental efforts
to explore coherence effects in quantum thermodynamics.

\section*{Acknowledgments}

We thank C. Ivan Henao and M. Herrera for very helpful discussions
concerning the relation between quantum friction and the generation
energy coherence. We acknowledge financial support from UFABC, CNPq,
CAPES, and FAPESP. This research was performed as part of the Brazilian
National Institute of Science and Technology for Quantum Information
(INCT-IQ). P. A. C. acknowledges CAPES and Templeton World Charity
Foundation, Inc. This publication was made possible through the support
of Grant No. TWCF0338 from Templeton World Charity Foundation, Inc.
The opinions expressed in this publication are those of the author(s)
and do not necessarily reflect the views of Templeton World Charity
Foundation, Inc.

\setcounter{section}{0} 
\global\long\def\thesection{Appendix \Alph{section}}%

\section{The Engine Cycle\label{sec:The-Engine-Cycle-1}}

\setcounter{figure}{0} \setcounter{equation}{0} 
\global\long\def\thefigure{A\arabic{figure}}%
 
\global\long\def\theequation{A\arabic{equation}}%

In Sec.~\ref{sec:The-Engine-Cycle} we explained the QOHE cycle,
however some important aspects were not thoroughly discussed. First,
we show that the energy transition probability of the expansion stroke
is the same as the energy transition probability of the compression
stroke. Then, we discuss the relation between the expansion and compression
driving fields, which are not the backward protocol of one another as considered
in some papers~\cite{Peterson2018}. Also, we show the expressions
of the master equation for incomplete thermalization.

\subsection*{Relation between expansion and compression strokes\label{subsec:Relation-between-expansion and compression strokes}}

\begin{figure}
\includegraphics[scale=0.71]{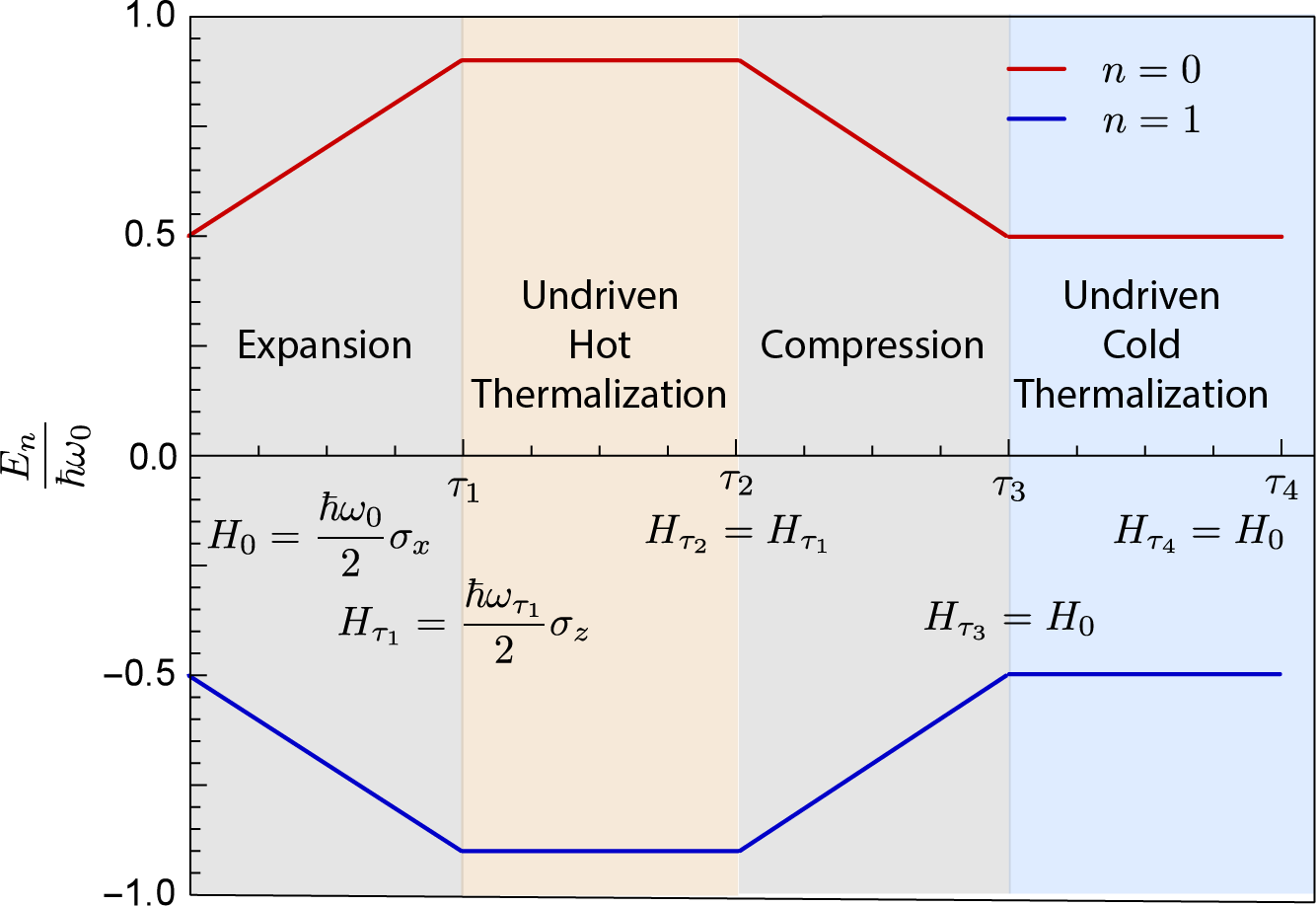}

\caption{Change in the eigenenergies during one cycle. For $n=0$ \textcolor{black}{[red upper
line] and $n=1$ [blue lower line]} the curves show the value of the instantaneous
eigenenergies during one cycle. From the parameters considered in
Sec.~\ref{sec:Thermodynamic-Quantities-and}, the relation between
the initial and final frequencies is $\omega_{\tau_{1}}=1.8\times\omega_{0}$.
With this information, we plotted the renormalized instantaneous eigenenergies
$E_{n}\left(t\right)/\hbar\omega_{0}$ as a function of time. Additionally, we
depicted each stroke as well as represented the Hamiltonian at the
four key instants of time. \label{fig:eigenenergies}}
\end{figure}

In Fig.~\ref{fig:eigenenergies} we show how the instantaneous eigenenergies
change during one cycle. The Hamiltonian of the total engine is given
by 
\begin{equation}
H\left(t\right)=\begin{cases}
H^{\text{exp}}\left(t\right) & t\in\left[0,\tau_{1}\right]\\
H^{\text{hot}}\left(t\right) & t\in\left[\tau_{1},\tau_{2}\right]\\
H^{\text{com}}\left(t\right) & t\in\left[\tau_{2},\tau_{3}\right]\\
H^{\text{cold}}\left(t\right) & t\in\left[\tau_{3},\tau_{4}\right],
\end{cases}
\end{equation}
where each of these Hamiltonians has been defined in the main text
{[}see Eq.~(\ref{eq:driving hamiltonian}){]}.

The unitary evolutions of the first and third strokes are 
\begin{equation}
U_{\tau_{1},0}=\exp\left\{ -\frac{i}{\hbar}\int_{0}^{\tau_{1}}dt\,H^{\text{exp}}\left(t\right)\right\} \label{eq:unitary expansion stroke}
\end{equation}
and 
\begin{equation}
V_{\tau_{3},\tau_{2}}=\exp\left\{ -\frac{i}{\hbar}\int_{\tau_{2}}^{\tau_{3}}dt\,H^{\text{com}}\left(t\right)\right\} ,
\end{equation}
respectively, where $H^{\text{exp}}\left(t\right)$ is defined in
the time interval $t\in\left[0,\tau_{1}\right]$ and $H^{\text{com}}\left(t\right)$
is defined in the time interval $t\in\left[\tau_{2},\tau_{3}\right]$.
Recall that $H^{\text{com}}\left(t\right)=H^{\text{exp}}\left(\tau_{3}-t\right)$,
where $\tau_{3}-\tau_{2}=\tau_{1}$, i.e., the expansion and compression
strokes take the same amount of time to be performed. Changing variables
in Eq.~(\ref{eq:unitary expansion stroke}) to $\tau_{1}+\tau_{2}-t=\tau_{3}-t$
one obtains 
\begin{align}
U_{\tau_{1},0}= & \exp\left\{ -\frac{i}{\hbar}\int_{\tau_{3}}^{\tau_{2}}\left(-dt\right)\,H^{\text{exp}}\left(\tau_{3}-t\right)\right\} \nonumber \\
= & \exp\left\{ -\frac{i}{\hbar}\int_{\tau_{2}}^{\tau_{3}}dt\,H^{\text{com}}\left(t\right)\right\} =V_{\tau_{3},\tau_{2}}.\label{eq:equivalence unitaries}
\end{align}
This seems a quite strange result. Even though the variation of the
Hamiltonian is different, the time-evolution operator coincides. However,
if we want to establish the physical condition $H^{\text{com}}\left(t\right)=H^{\text{exp}}\left(\tau_{3}-t\right)$
that makes the third-stroke driving Hamiltonian go back to the initial
Hamiltonian of the first stroke passing through the same Hamiltonians
in between, Eq.~(\ref{eq:equivalence unitaries}) is true as
demonstrated above.

The definitions of the transition probabilities between energy states
of the expansion and compression strokes are 
\begin{equation}
p_{\tau_{1},0}^{\text{exp}}\left(m|n\right)=\left|\Bra{E_{m}^{\tau_{1}}}U_{\tau_{1},0}\Ket{E_{n}^{0}}\right|^{2}
\end{equation}
and 
\begin{equation}
p_{\tau_{3},\tau_{2}}^{\text{com}}\left(m|n\right)=\left|\Bra{E_{m}^{0}}V_{\tau_{3},\tau_{2}}\Ket{E_{n}^{\tau_{1}}}\right|^{2}.
\end{equation}
By definition, the energy transition probability of the expansion
and compression strokes are 
\begin{equation}
\xi\left(\tau_{1},0\right)=\left|\Bra{E_{1}^{\tau_{1}}}U_{\tau_{1},0}\Ket{E_{0}^{0}}\right|^{2}=\left|\Bra{E_{1}^{\tau_{1}}}U_{\tau_{1},0}\Ket{E_{0}^{0}}\right|^{2}
\end{equation}
and 
\begin{equation}
\zeta\left(\tau_{3},\tau_{2}\right)=\left|\Bra{E_{1}^{0}}V_{\tau_{3},\tau_{2}}\Ket{E_{0}^{\tau_{1}}}\right|^{2},
\end{equation}
respectively. Using $V_{\tau_{3},\tau_{2}}=U_{\tau_{1},0}$ from Eq.~(\ref{eq:equivalence unitaries})
and opening the modulus square one can easily show that 
\begin{equation}
\zeta\left(\tau_{3},\tau_{2}\right)=\xi\left(\tau_{1},0\right).
\end{equation}

\subsection*{Incomplete thermalization relations}

Suppose the initial state of a qubit $\rho_{0}$ is given by the Bloch
vector $\mathbf{r}_{0}=\left(r_{x}\left(0\right),r_{y}\left(0\right),r_{z}\left(0\right)\right)$,
where $r_{i}\left(t\right)=\text{Tr}\left[\sigma_{i}\rho_{t}\right]$
are the instantaneous Bloch components for $i=x,y,z$. From Refs.~\cite{Breuerbook2003,Chakraborty2016},
the master equation of a qubit interacting with a Markovian heat reservoir
and where the Hamiltonian is fixed at $H=\frac{\hbar\omega}{2}\sigma_{z}$
is given by Eq.~(\ref{eq:master equation qubit}). The solutions of
the Bloch vector components at the end of the second stroke are given
by 
\begin{equation}
r_{x}\left(\tau_{\text{therm}}^{\text{h}}\right)=r_{x}\left(\tau_{1}\right)e^{-\gamma\tau_{\text{therm}}^{\text{h}}/2}
\end{equation}
\begin{equation}
r_{y}\left(\tau_{\text{therm}}^{\text{h}}\right)=r_{y}\left(\tau_{1}\right)e^{-\gamma\tau_{\text{therm}}^{\text{h}}/2}
\end{equation}
\begin{equation}
r_{z}\left(\tau_{\text{therm}}^{\text{h}}\right)=\left(r_{z}\left(\tau_{1}\right)+g\right)e^{-\gamma\tau_{\text{therm}}^{\text{h}}}-g_{\text{H}},
\end{equation}
where $\gamma=\gamma_{0}\left(2N_{\text{BE}}+1\right)$, $g=\frac{\gamma_{0}}{\gamma}$,
and the remaining parameters have been defined in the main text. One
of the coherence elements at the end of the incomplete thermalization
oscillates as 
\begin{align}
\Bra{E_{1}^{\tau_{1}}}\rho_{\tau_{2}}\Ket{E_{0}^{\tau_{1}}}= & \Bra{0}\rho_{\tau_{2}}\Ket{1}=\frac{r_{x}\left(\tau_{\text{therm}}^{\text{h}}\right)-ir_{y}\left(\tau_{\text{therm}}^{\text{h}}\right)}{2}\nonumber \\
= & \Bra{E_{1}^{\tau_{1}}}\rho_{\tau_{1}}\Ket{E_{0}^{\tau_{1}}}e^{+i\tau_{\text{therm}}^{\text{h}}\omega_{\tau_{1}}}e^{-\gamma\tau_{\text{therm}}^{\text{h}}/2}\label{eq:34}
\end{align}
These oscillatory terms are the origins of the oscillations in Figs.~\ref{fig:Thermodynamic-Quantities.}\textbf{(c)}
and $\mathbf{(d)}$.

\setcounter{figure}{0} \setcounter{equation}{0} 
\global\long\def\thefigure{B\arabic{figure}}%
 
\global\long\def\theequation{B\arabic{equation}}%

\section{Entropy Production\label{subsec:The-thermal-efficiency}}

\subsection*{Efficiency and entropy production}

Since the working substance ends up in the same initial state as the
engine cycle due to the complete thermalization with the cold heat
reservoir, the total change in energy and entropy are 
\begin{equation}
\left\langle W_{1}\right\rangle +\left\langle W_{3}\right\rangle +\left\langle Q_{\text{h}}\right\rangle +\left\langle Q_{\text{c}}\right\rangle =0\label{eq:energy conservation}
\end{equation}
and 
\begin{equation}
\Delta S_{2}+\Delta S_{4}=0,\label{eq:entropy conservation}
\end{equation}
respectively. In Eq.~(\ref{eq:entropy conservation}), $\Delta S_{4}=S\left(\rho_{0}^{\text{eq,c}}\right)-S\left(\rho_{\tau_{3}}\right)$
and $\Delta S_{2}=S\left(\rho_{\tau_{2}}\right)-S\left(\rho_{\tau_{1}}\right)$
are the changes in entropy during the fourth and second strokes, respectively.
The first and third strokes are unitary and hence do not change the
entropy. These equations express the conservation of energy and entropy
in one cycle.

The total change of entropy of the working substance is equal to the
total entropy production plus the total heat flux. Since the working
substance interacts with two heat reservoirs, it contains two contributions
to the heat flux, namely, $\beta_{\text{c}}\left\langle Q_{\text{c}}\right\rangle $
and $\beta_{\text{h}}\left\langle Q_{\text{h}}\right\rangle $. Therefore,
\begin{equation}
\Delta S_{\text{total}}=\left\langle \Sigma_{\text{total}}\right\rangle +\beta_{\text{h}}\left\langle Q_{\text{h}}\right\rangle +\beta_{\text{c}}\left\langle Q_{\text{c}}\right\rangle .\label{eq:auxiliary entropy}
\end{equation}
From entropy conservation $\Delta S_{\text{total}}=0$, implying that
the total amount of entropy produced is dispersed to the reservoirs.

The engine efficiency is given by 
\begin{equation}
\eta=\frac{-\left\langle W_{1}\right\rangle -\left\langle W_{3}\right\rangle }{\left\langle Q_{\text{h}}\right\rangle }=1+\frac{\beta_{\text{c}}\left\langle Q_{\text{c}}\right\rangle }{\beta_{\text{c}}\left\langle Q_{\text{h}}\right\rangle },\label{eq:auxiliar efficiency}
\end{equation}
where Eq.~(\ref{eq:energy conservation}) has been employed and $\beta_{\text{c}}$
has been conveniently introduced. Substituting $\beta_{\text{c}}\left\langle Q_{\text{c}}\right\rangle $
from Eq.~(\ref{eq:auxiliary entropy}) into Eq.~(\ref{eq:auxiliar efficiency})
and rearranging the terms one obtains 
\begin{equation}
\eta=\eta_{\text{Carnot}}-\frac{\left\langle \Sigma_{\text{total}}\right\rangle }{\beta_{\text{c}}\left\langle Q_{\text{h}}\right\rangle }\label{eq:auxiliary efficiency}
\end{equation}
which is Eq.~(\ref{eq:efficiency carnot}).

\subsection*{Entropy production and thermal divergences}

In this subsection, we derive the expression for the total entropy production
as a function of the thermal divergences {[}Eq.~(\ref{eq:entropy production total}){]}.
We begin with Eq.~(\ref{eq:auxiliar efficiency}) 
\begin{equation}
\eta=1+\frac{\left\langle Q_{\text{c}}\right\rangle }{\left\langle Q_{\text{h}}\right\rangle }=1+\frac{\mathcal{E}_{0}-\mathcal{E}_{\tau_{3}}}{\left\langle Q_{\text{h}}\right\rangle },\label{eq:41}
\end{equation}
where in the last equality we wrote the cold heat in terms of the
internal energies.

Next, we use the following relation valid for the thermal divergence
(see the Supplemental Material of Ref.~\cite{Camati2016} for a quick
derivation). Let $\rho_{t}$ be an arbitrary state in some time $t$
with Hamiltonian $H\left(t\right)$ and $\rho_{t}^{\text{eq}}$ an
associated Gibbs state with same Hamiltonian and some reference inverse
temperature $\beta$; then 
\begin{equation}
D\left(\rho_{t}||\rho_{t}^{\text{eq}}\right)=\beta\left[\mathcal{E}\left(\rho_{t}\right)-F_{t}^{\text{eq}}\right]-S\left(\rho_{t}\right),\label{eq:fundamental relation}
\end{equation}
where $F_{t}^{\text{eq}}=-\left(\beta\right)^{-1}\ln Z_{t}$ is the
associated free energy and $Z_{t}=\text{Tr}\left[e^{-\beta H_{t}}\right]$
is the associated partition function.

We substitute the internal energies in the efficiency using the expressions
\begin{equation}
\beta_{\text{c}}\mathcal{E}\left(\rho_{0}^{\text{eq,c}}\right)=D\left(\rho_{0}^{\text{eq,c}}||\rho_{0}^{\text{eq,c}}\right)+S\left(\rho_{0}^{\text{eq,c}}\right)+\beta_{\text{c}}F_{0}^{\text{eq,c}}
\end{equation}
and 
\begin{equation}
\beta_{\text{c}}\mathcal{E}\left(\rho_{\tau_{3}}\right)=D\left(\rho_{\tau_{3}}||\rho_{0}^{\text{eq,c}}\right)+S\left(\rho_{\tau_{3}}\right)+\beta_{\text{c}}F_{0}^{\text{eq,c}},
\end{equation}
where we have used $\rho_{\tau_{3}}^{\text{eq,c}}=\rho_{0}^{\text{eq,c}}$,
and $F_{\tau_{3}}^{\text{eq,c}}=F_{0}^{\text{eq,c}}$ because the
Hamiltonian is the same at times $\tau_{3}$ and zero (see Fig.~\ref{fig:eigenenergies}).
Hence, 
\begin{equation}
\eta=1+\frac{\Delta S_{4}-D\left(\rho_{\tau_{3}}||\rho_{0}^{\text{eq,c}}\right)}{\beta_{\text{c}}\left\langle Q_{\text{h}}\right\rangle }=1+\frac{-\Delta S_{2}-D\left(\rho_{\tau_{3}}||\rho_{0}^{\text{eq,c}}\right)}{\beta_{\text{c}}\left\langle Q_{\text{h}}\right\rangle },\label{eq:45}
\end{equation}
where we already canceled the free-energy terms and used Eq.~(\ref{eq:entropy conservation}).
Using Eq.~(\ref{eq:fundamental relation}), we substitute the von
Neumann entropies 
\begin{equation}
S\left(\rho_{\tau_{1}}\right)=\beta_{\text{h}}\mathcal{E}\left(\rho_{\tau_{1}}\right)-\beta_{\text{h}}F_{\tau_{1}}^{\text{eq,h}}-D\left(\rho_{\tau_{1}}||\rho_{\tau_{1}}^{\text{eq,h}}\right)
\end{equation}
and 
\begin{equation}
S\left(\rho_{\tau_{2}}\right)=\beta_{\text{h}}\mathcal{E}\left(\rho_{\tau_{2}}\right)-\beta_{\text{h}}F_{\tau_{1}}^{\text{eq,h}}-D\left(\rho_{\tau_{2}}||\rho_{\tau_{1}}^{\text{eq,h}}\right)
\end{equation}
in order to obtain 
\begin{equation}
\Delta S_{2}=\beta_{\text{h}}\left\langle Q_{\text{h}}\right\rangle -D\left(\rho_{\tau_{2}}||\rho_{\tau_{1}}^{\text{eq,h}}\right)+D\left(\rho_{\tau_{1}}||\rho_{\tau_{1}}^{\text{eq,h}}\right),\label{eq: 48}
\end{equation}
where we already used the fact that $\rho_{\tau_{2}}^{\text{eq,h}}=\rho_{\tau_{1}}^{\text{eq,h}}$,
$F_{\tau_{2}}^{\text{eq,h}}=F_{\tau_{1}}^{\text{eq,h}}$, because
the Hamiltonians are the same at times $\tau_{1}$ and $\tau_{2}$.
Replacing Eq.~(\ref{eq: 48}) into Eq.~(\ref{eq:45}) and rearranging
the terms we obtain 
\begin{equation}
\eta=\eta_{\text{Carnot}}-\frac{D\left(\rho_{\tau_{1}}||\rho_{\tau_{1}}^{\text{eq,h}}\right)-D\left(\rho_{\tau_{2}}||\rho_{\tau_{1}}^{\text{eq,h}}\right)+D\left(\rho_{\tau_{3}}||\rho_{0}^{\text{eq,c}}\right)}{\beta_{\text{c}}\left\langle Q_{\text{h}}\right\rangle }.
\end{equation}
Comparing this equation to the efficiency in Eq.~(\ref{eq:auxiliary efficiency}),
one obtains 
\begin{equation}
\left\langle \Sigma_{\text{total}}\right\rangle =D\left(\rho_{\tau_{1}}||\rho_{\tau_{1}}^{\text{eq,h}}\right)-D\left(\rho_{\tau_{2}}||\rho_{\tau_{1}}^{\text{eq,h}}\right)+D\left(\rho_{\tau_{3}}||\rho_{0}^{\text{eq,c}}\right),
\end{equation}
which is Eq. (\ref{eq:entropy production total}).

\setcounter{figure}{0} \setcounter{equation}{0} 
\global\long\def\thefigure{C\arabic{figure}}%
 
\global\long\def\theequation{C\arabic{equation}}%

\section{The quasistatic divergences\label{sec:The-quasistatic-divergences}}

Before we demonstrate the expression for the quantum friction $\mathcal{F}$
we need to obtain an expression for the quasistatic divergence similar
to Eq.~(\ref{eq:fundamental relation}) for the thermal divergence.

In the first stroke, the initial state is always the cold Gibbs state
$\rho_{0}^{\text{eq,c}}=e^{-\beta_{\text{c}}H_{0}}/Z_{0}^{\text{c}}=\sum_{n}p_{n}^{\text{eq,c},0}\Ket{E_{n}^{0}}\Bra{E_{n}^{0}}$,
where $p_{n}^{\text{eq,c},0}$ are the respective Boltzmann weights.
The the reference state for the first-stroke quasistatic divergence
is 
\begin{equation}
\rho_{\tau_{1}}^{\text{qs,c}}=\sum_{n}p_{n}^{\text{eq,c},0}\Ket{E_{n}^{\tau_{1}}}\Bra{E_{n}^{\tau_{1}}},
\end{equation}
where the eigenstates changed without changing the populations of
the state. The quasistatic divergence is given by 
\begin{equation}
D\left(\rho_{\tau_{1}}||\rho_{\tau_{1}}^{\text{qs,c}}\right)=-\text{Tr}\left[\rho_{\tau_{1}}\ln\rho_{\tau_{1}}^{\text{qs,c}}\right]-S\left(\rho_{\tau_{1}}\right).
\end{equation}
Expanding the trace in the basis of $H_{\tau_{1}}$ and using $\ln p_{n}^{\text{eq,c},0}=-\beta_{\text{c}}\left(E_{n}^{0}-F_{0}^{\text{eq,c}}\right)$
the first term of the divergence can be written as 
\begin{equation}
-\text{Tr}\left[\rho_{\tau_{1}}\ln\rho_{\tau_{1}}^{\text{qs,c}}\right]=\sum_{n}\beta_{\text{c}}E_{n}^{0}\text{Tr}\left[\rho_{\tau_{1}}\Ket{E_{n}^{\tau_{1}}}\Bra{E_{n}^{\tau_{1}}}\right]-\beta_{\text{c}}F_{0}^{\text{eq,c}}.
\end{equation}

Now comes an important assumption that constrains the derivation.
We assume that the ratio $E_{n}^{\tau_{1}}/E_{n}^{0}$ for every $n$
between the final and initial energies is constant. This applies at
least for a qubit or a harmonic oscillator. In our case, we are considering
a qubit working substance, so this ratio is $E_{n}^{\tau_{1}}/E_{n}^{0}=\omega_{\tau_{1}}/\omega_{0}$
for $n=0,1$. We multiply the term $E_{n}^{0}$ by $E_{n}^{\tau_{1}}/E_{n}^{\tau_{1}}=1$
and rearrange the eigenenergies $E_{n}^{\tau_{1}}$ inside the trace,
obtaining the spectral decomposition of the final Hamiltonian $H_{\tau_{1}}=\sum_{n}E_{n}^{0}\Ket{E_{n}^{\tau_{1}}}\Bra{E_{n}^{\tau_{1}}}$.
Therefore, we obtain the following expression for quasistatic divergence
\begin{equation}
D\left(\rho_{\tau_{1}}||\rho_{\tau_{1}}^{\text{qs,c}}\right)=\beta_{\text{c}}\frac{\omega_{0}}{\omega_{\tau_{1}}}\mathcal{E}\left(\rho_{\tau_{1}}\right)-\beta_{\text{c}}F_{0}^{\text{eq,c}}-S\left(\rho_{\tau_{1}}\right),\label{eq:53}
\end{equation}
which is the relation we were seeking.

Now, we want to obtain the same relation for the third stroke. However,
the initial state is not the Gibbs state, in general, since an incomplete
thermalization is performed. As we mentioned in the main text, the
quasistatic state that would be obtained if we performed the compression
stroke quasistatically is not the reference state used in the quasistatic
divergence. Instead, we use the reference state $\rho_{\tau_{3}}^{\text{qs,h}}$,
which is the quasistatic state that would have been obtained if the
transformation was performed quasistatically and the initial state
was the hot Gibbs state $\rho_{\tau_{2}}^{\text{eq,h}}=\sum_{n}p_{n}^{\text{eq,h},\tau_{1}}\Ket{E_{n}^{\tau_{1}}}\Bra{E_{n}^{\tau_{1}}}$.
Hence, 
\begin{equation}
\rho_{\tau_{3}}^{\text{qs,h}}=\sum_{n}p_{n}^{\text{eq,h},\tau_{1}}\Ket{E_{n}^{0}}\Bra{E_{n}^{0}}.
\end{equation}
The quasistatic divergence we use is 
\begin{equation}
D\left(\rho_{\tau_{3}}||\rho_{\tau_{3}}^{\text{qs,h}}\right)=-\text{Tr}\left[\rho_{\tau_{3}}\ln\rho_{\tau_{3}}^{\text{qs,h}}\right]-S\left(\rho_{\tau_{3}}\right).\label{eq:55}
\end{equation}
Using the same strategy as before to calculate the first term we obtain
\begin{equation}
-\text{Tr}\left[\rho_{\tau_{3}}\ln\rho_{\tau_{3}}^{\text{qs,h}}\right]=\beta_{\text{h}}\frac{\omega_{\tau_{1}}}{\omega_{0}}\mathcal{E}\left(\rho_{\tau_{3}}\right)-\beta_{\text{h}}F_{\tau_{1}}^{\text{eq,h}},
\end{equation}
where we used $\ln p_{n}^{\text{eq,h},\tau_{1}}=-\beta_{\text{h}}\left(E_{n}^{\tau_{1}}-F_{\tau_{1}}^{\text{eq,h}}\right)$
and the spectral decomposition of the initial Hamiltonian is $H_{0}=\sum_{n}E_{n}^{0}\Ket{E_{n}^{0}}\Bra{E_{n}^{0}}$.
Therefore, the relation we seek is 
\begin{equation}
D\left(\rho_{\tau_{3}}||\rho_{\tau_{3}}^{\text{qs,h}}\right)=\beta_{\text{h}}\frac{\omega_{\tau_{1}}}{\omega_{0}}\mathcal{E}\left(\rho_{\tau_{3}}\right)-\beta_{\text{h}}F_{\tau_{1}}^{\text{eq,h}}-S\left(\rho_{\tau_{3}}\right).\label{eq:57}
\end{equation}

\setcounter{figure}{0} \setcounter{equation}{0} 
\global\long\def\thefigure{D\arabic{figure}}%
 
\global\long\def\theequation{D\arabic{equation}}%

\section{Quantum Friction\label{sec:The-quasistatic-efficiency lag}}

Using Eqs.~(\ref{eq:53}) and (\ref{eq:57}), we can derive the expression
for the quantum friction. We begin the derivation from Eq.~(\ref{eq:41})
for the efficiency 
\begin{equation}
\eta=1+\frac{\beta_{\text{c}}\mathcal{E}_{0}-\beta_{\text{c}}\mathcal{E}_{\tau_{3}}}{\beta_{\text{c}}\left\langle Q_{\text{h}}\right\rangle },\label{eq:58}
\end{equation}
where we multiplied the second term by $1=\beta_{\text{c}}/\beta_{\text{c}}$.
Let us consider the numerator in the second term. We want to relate
the initial energy to the quasistatic divergence. We use the expression
of the quasistatic divergence given by Eq.~(\ref{eq:53}), $S\left(\rho_{\tau_{1}}\right)=S\left(\rho_{0}^{\text{eq,c}}\right)$
because the first stroke is unitary, and the identity $S\left(\rho_{0}^{\text{eq,c}}\right)=\beta_{\text{c}}\mathcal{E}\left(\rho_{0}^{\text{eq,c}}\right)-\beta_{\text{c}}F_{0}^{\text{eq,c}}$.
Therefore, we can obtain the relation 
\begin{equation}
\beta_{\text{c}}\mathcal{E}\left(\rho_{0}^{\text{eq,c}}\right)=\beta_{\text{c}}\frac{\omega_{0}}{\omega_{\tau_{1}}}\mathcal{E}\left(\rho_{\tau_{1}}\right)-D\left(\rho_{\tau_{1}}||\rho_{\tau_{1}}^{\text{qs,c}}\right),\label{eq:59}
\end{equation}
where we have isolated the initial energy. Equation~(\ref{eq:57})
already relates the internal energy $\mathcal{E}_{\tau_{3}}$ to the
quasistatic divergence. However, to derive the desired expression
we must eliminate the von Neumann entropy $S\left(\rho_{\tau_{3}}\right)$
from the equation. We do this by first using $S\left(\rho_{\tau_{3}}\right)=S\left(\rho_{\tau_{2}}\right)$,
since the third stroke is unitary, and then using the relation for
the thermal divergence Eq.~(\ref{eq:fundamental relation}). Hence,
we obtain 
\begin{align}
\beta_{\text{c}}\mathcal{E}\left(\rho_{\tau_{3}}\right)= & \frac{\beta_{\text{c}}\omega_{0}}{\beta_{\text{h}}\omega_{\tau_{1}}}\left[D\left(\rho_{\tau_{3}}||\rho_{\tau_{3}}^{\text{qs,h}}\right)-D\left(\rho_{\tau_{2}}||\rho_{\tau_{1}}^{\text{eq,h}}\right)\right]\nonumber \\
 & +\beta_{\text{c}}\frac{\omega_{0}}{\omega_{\tau_{1}}}\mathcal{E}\left(\rho_{\tau_{2}}\right),\label{eq:60}
\end{align}
where we have already isolated the internal energy $\mathcal{E}_{\tau_{3}}$.
Substituting Eqs.~(\ref{eq:59}) and (\ref{eq:60}) into Eq.~(\ref{eq:58}),
and manipulating the terms we arrive at 
\begin{equation}
\eta=\eta_{\text{Otto}}-\frac{\mathcal{F}}{\beta_{\text{c}}\left\langle Q_{\text{h}}\right\rangle },
\end{equation}
which is Eq.~(\ref{eq:efficiency otto}), and we identify 
\begin{equation}
\mathcal{F}=D\left(\rho_{\tau_{1}}||\rho_{\tau_{1}}^{\text{qs,c}}\right)+\frac{\omega_{0}\beta_{\text{c}}}{\omega_{\tau_{1}}\beta_{\text{h}}}\left[D\left(\rho_{\tau_{3}}||\rho_{\tau_{3}}^{\text{qs,h}}\right)-D\left(\rho_{\tau_{2}}||\rho_{\tau_{2}}^{\text{eq,h}}\right)\right]
\end{equation}
as the quantum friction {[}Eq.~(\ref{eq:quantum friction lag}){]}.

\setcounter{figure}{0} \setcounter{equation}{0} 
\global\long\def\thefigure{E\arabic{figure}}%
 
\global\long\def\theequation{E\arabic{equation}}%

\section{Interference Energy Contribution\label{sec:Transferred-coherence-energy-contribution}}

In this appendix, we demonstrate the energy contribution due to the
interference effect given by Eq. (\ref{eq:energy trans}). This energy
contribution comes, in fact, from the internal energy at $t=\tau_{3}$.
This internal energy is given by $\mathcal{E}_{\tau_{3}}=\mathcal{E}\left(\rho_{\tau_{3}}\right)=\text{Tr}\left[\rho_{\tau_{3}}H_{0}\right]$.
Let $\rho_{\tau_{2}}=\sum_{nm}\rho_{nm}\left(\tau_{2}\right)\Ket{E_{n}^{\tau_{1}}}\Bra{E_{m}^{\tau_{1}}}$
denote the state at the end of the second stroke, where $\rho_{nm}\left(\tau_{2}\right)=\Bra{E_{n}^{\tau_{1}}}\rho_{\tau_{2}}\Ket{E_{m}^{\tau_{1}}}$.
The term $\rho_{10}\left(\tau_{2}\right)$ is given by Eq.~(\ref{eq:34}).
The internal energy can be decomposed as 
\begin{equation}
\mathcal{E}_{\tau_{3}}=\sum_{mm'}\rho_{mm'}\left(\tau_{2}\right)\sum_{n}E_{n}^{0}a_{nm}^{\text{com}}\left(\tau_{3},\tau_{2}\right)a_{nm'}^{\text{com}*}\left(\tau_{3},\tau_{2}\right),
\end{equation}
where we used the spectral decomposition of the initial Hamiltonian,
used $\rho_{\tau_{3}}=V_{\tau_{3},\tau_{2}}\rho_{\tau_{2}}V_{\tau_{3},\tau_{2}}^{\dagger}$,
opened the trace, and used the definition of the energy transition
amplitude defined in the main text. Splitting the summation $\sum_{m,m'}=\sum_{m}+\sum_{m\neq m'}$,
the second term will be the interference contribution 
\begin{equation}
\mathcal{E}_{\text{trans}}=\sum_{m\neq m'}\rho_{mm'}\left(\tau_{2}\right)\sum_{n}E_{n}^{0}a_{nm}^{\text{com}}\left(\tau_{3},\tau_{2}\right)a_{nm'}^{\text{com}*}\left(\tau_{3},\tau_{2}\right).
\end{equation}
Writing the sum over $m$ and $m'$ explicitly, one finds that the
terms are the complex conjugates of each other. Hence, 
\begin{equation}
\mathcal{E}_{\text{trans}}=2\sum_{n}E_{n}^{0}\text{Re}\left\{ \rho_{10}\left(\tau_{2}\right)a_{n1}^{\text{com}}\left(\tau_{3},\tau_{2}\right)a_{n0}^{\text{com}*}\left(\tau_{3},\tau_{2}\right)\right\} .
\end{equation}
Using Eq.~(\ref{eq:34}) we finally arrive at 
\begin{align}
\mathcal{E}_{\text{trans}}= & 2\sum_{n}E_{n}^{0}e^{-\gamma\tau_{\text{therm}}^{\text{h}}/2}\nonumber \\
 & \times\text{Re}\left\{ \rho_{10}\left(\tau_{1}\right)e^{+i\tau_{2}\omega_{\tau_{1}}\tau_{\text{therm}}^{\text{h}}}a_{n0}^{\text{com}}a_{n1}^{\text{com}*}\right\} ,
\end{align}
where $\rho_{10}\left(\tau_{1}\right)=\Bra{E_{1}^{\tau_{1}}}\rho_{\tau_{1}}\Ket{E_{0}^{\tau_{1}}}$.

\end{document}